\documentclass[a4paper,11pt]{article}
\pdfoutput=1 

\usepackage{jheppub} 

\usepackage[T1]{fontenc} 
\usepackage{comment}
\usepackage{slashed}
\usepackage{graphicx}

\preprint{
  {\small
    \hphantom{.}\hfill HIP-2021-18/TH\\
    \hphantom{.}\hfill IPPP/20/100\\
    \hphantom{.}\hfill LTH 1260\\
  }
}

\title{\boldmath Exclusive heavy vector meson electroproduction to NLO in collinear factorisation }

\author[a,b,c]{C.~A.~Flett,}
\author[c]{J.~A.~Gracey,}
\author[d]{S.~P.~Jones}
\author[c]{and T.~Teubner}

\affiliation[a]{Department of Physics, University of Jyv\"{a}skyl\"{a}, P.O. Box 35, 40014 University of Jyv\"{a}skyl\"{a}, Finland}
\affiliation[b]{Helsinki Institute of Physics, P.O. Box 64, 00014 University of Helsinki, Finland}
\affiliation[c]{Department of Mathematical Sciences, University of Liverpool, Liverpool, L69 3BX, U.K.}
\affiliation[d]{Institute for Particle Physics Phenomenology, Durham University, Durham, DH1 3LE, U.K.}

\emailAdd{chris.a.flett@jyu.fi}
\emailAdd{gracey@liverpool.ac.uk}
\emailAdd{stephen.jones@durham.ac.uk}
\emailAdd{thomas.teubner@liverpool.ac.uk}

\abstract{We compute the exclusive electroproduction, $\gamma^* p \rightarrow V p$, of heavy quarkonia~$V$ to NLO in the collinear factorisation scheme, which has been formally proven for this process. The inclusion of an off-shell virtuality $Q^2$ carried by the photon extends the photoproduction phase space of the exclusive heavy quarkonia observable to electroproduction kinematics. This process is relevant for diffractive scattering at HERA and the upcoming EIC, as well as at the proposed LHeC and FCC.}

\begin{document} 
\maketitle
\flushbottom

\section{Introduction}
\label{sec:intro}
The exclusive production of vector mesons has long been an interesting and attractive observable to study. First measured in the fixed target mode and then in diffractive deep inelastic scattering~(DIS) events at the $ep$ linear HERA collider more than 25 years ago, they constitute $\sim 10\%$ of the total inclusive DIS cross-section and are characterised by the presence of a large rapidity gap. They provide a means to investigate the phenomenology of quarkonium production and function as more sensitive probes of the low-$x$ and low scale input gluon parton distribution than any other known high-energy physics phenomenon. 

In 1993, around the same time as the first measurements of such diffractive activity in a collider environment, the exclusive {\it electro}production of a {\it heavy} vector meson~(HVM), $V=J/\psi, \Upsilon, \psi(2S),\dots$ via $\gamma^* p \rightarrow Vp$, was showcased to be proportional to the square of the gluon parton distribution function (PDF)~\cite{Ryskin:1992ui} in the leading logarithmic approximation~(LLA) of perturbative QCD~(pQCD), within the framework of $k_T$-factorisation.  See~\cite{Ivanov:2004ax} for a review. This coincides with the leading-order (LO) term in the collinear factorisation scheme, which we use here. This process is on solid ground thanks to a proven factorisation theorem~\cite{Collins:1996fb} and the pQCD treatment is justified for large virtualities $Q^2$ of the photon, $\gamma^*$. 

On the experimental side, the HERMES collaboration~\cite{Augustyniak:2013ypa} reported leptoproduction measurements for the lightest vector mesons in the range $1~\text{GeV}^2 < Q^2 < 7~\text{GeV}^2$ in fixed-target kinematics. Exclusive electroproduction data for the $J/\psi$ HVM via dimuon and dielectron decays has been measured in the collider mode at HERA in a narrow range of photon virtualities at both ZEUS and H1 experiments, extending up to the largest bin of $\langle Q^2 \rangle = 22.4~\text{GeV}^2$~\cite{Chekanov:2004mw,Aktas:2005xu}. As in photoproduction, the cross section exhibits a steep rise with increasing centre of mass energies of the $\gamma^* p \rightarrow J/\psi p$ subprocess. Today, in the LHC era of collider physics, central exclusive {\it photo}production of vector mesons $V$ have been measured in the forward rapidity interval $2.0 < Y < 4.5$ by the LHCb collaboration via ultraperipheral $pp \rightarrow p+V+p$ collisions instead~\cite{Aaij:2014iea,Aaij:2015kea,Aaij:2018arx}. These are driven by the hard scattering subprocesses $\gamma p \rightarrow Vp$, measured directly at HERA. Here, the photoproduction reaction~($Q^2 \approx 0$) is initiated by a real on-shell photon, $\gamma$. Despite the near vanishing of this scale, the factorisation theorems are still assumed to hold for photoproduction since the masses of the produced final state heavy mesons are above the perturbative scale.

Various theoretical models within pQCD exist in the literature that provide a description of the exclusive heavy vector meson photo and electroproduction processes~\cite{Ivanov:2004ax}. In the colour dipole approach, the exclusive HVM formation is dominated by scatterings in which the photon fluctuates into a $q \bar q$ pair with a transverse separation $r \approx 0$, carrying fractions $z \approx 1/2$ and $1-z \approx 1/2$ of the incoming photon momentum. The dipole model formulation is also able to describe light meson production and photon hadron scattering and is equivalent to the $k_T$-factorisation formalism in the leading $\ln(1/x)$ approximation. Following earlier work, in~\cite{Jones:2016icr}, the explicit $k_T$ integral had been performed in the last cell of evolution, in effect leading to a description beyond the LLA, to mimic a subset of the full next-to-leading order~(NLO) contribution. This accounts for those terms in the conventional DGLAP evolution that are enhanced at small $x$ and that in an axial gauge correspond strictly to gluon-ladder-rung Feynman diagrams. However, we emphasise this does not encompass the complete NLO contribution that one would obtain in the conventional systematic evaluation of Feynman diagrams within the $\overline{\text{MS}}$ collinear factorisation framework. 

In this work, we remain entirely within the collinear factorisation set-up at NLO to extract the electroproduction renormalised transverse and longitudinal coefficient functions to NLO in the $\overline{\text{MS}}$ scheme. Previously in the literature~\cite{Ivanov:2004vd, stethesis}, next-to-leading order $\overline{\text{MS}}$ coefficient functions were calculated in the case of photoproduction of HVMs. The authors of~\cite{Ivanov:2004vd} constructed the imaginary part of Feynman diagrams via Cutkosky-cuts in the $s$-channel and then restored their real parts using the corresponding $u$-channel contributions, via a dispersion relation. In our approach for electroproduction, we directly compute the real and imaginary parts of the amplitude using an integral reduction procedure. As will be discussed and explained, the zero photon virtuality limit of our electroproduction coefficient functions should coincide with these photoproduction results. Note that a subset of the authors in~\cite{Ivanov:2004vd} also computed the electroproduction of {\it light} neutral vector mesons $\tilde{V} = \rho^0, \omega$ and $\phi$~\cite{Ivanov:2004pp}. In our computation, both the virtuality of the photon and the mass of the heavy quark constitute massive scales, which add complexity.

The work presented here is the first step in driving the prediction of exclusive heavy vector meson electroproduction to NLO. As such it represents the technical details of the
underlying field theory construction.
In the case of photoproduction further steps, including scale fixing and a $Q_0$ subtraction, were required to improve the convergence of the perturbative prediction before the results could be applied in phenomenological studies.
For exclusive $J/\psi$ photoproduction these steps were carried out in \cite{Jones:2015nna, Jones:2016ldq, Flett:2019pux, Flett:2020duk} and used to present predictions which described data from HERA and LHCb and to determine the gluon distribution at low $x$ and scale. Going forward, our electroproduction results will be useful for studying exclusive processes at the upcoming EIC~\cite{AbdulKhalek:2021gbh} and at the proposed LHeC~\cite{LHeC:2020van} and FCC~\cite{FCC:2018byv}, which we leave for future work. 

Our paper is organised as follows. In Section 2, we give our set-up and model assumptions within collinear factorisation at NLO. In Section 3, we outline the workflow of our calculation. In Section 4 and Section 5, we give analytic expressions for the LO and NLO coefficient functions for the quark and gluon initiated subprocesses. We finish, in Section 6, by checking the explicit cancellation of collinear divergences to NLO within a consistent UV and IR subtraction scheme, before making a comparison with literature in Section 7 and presenting our conclusions in Section 8.

\section{Notation and collinear factorisation}
\subsection{Kinematics and set-up}
\label{setup}
\begin{figure}[h!]
  \begin{center}
  \includegraphics[width=0.4\textwidth]{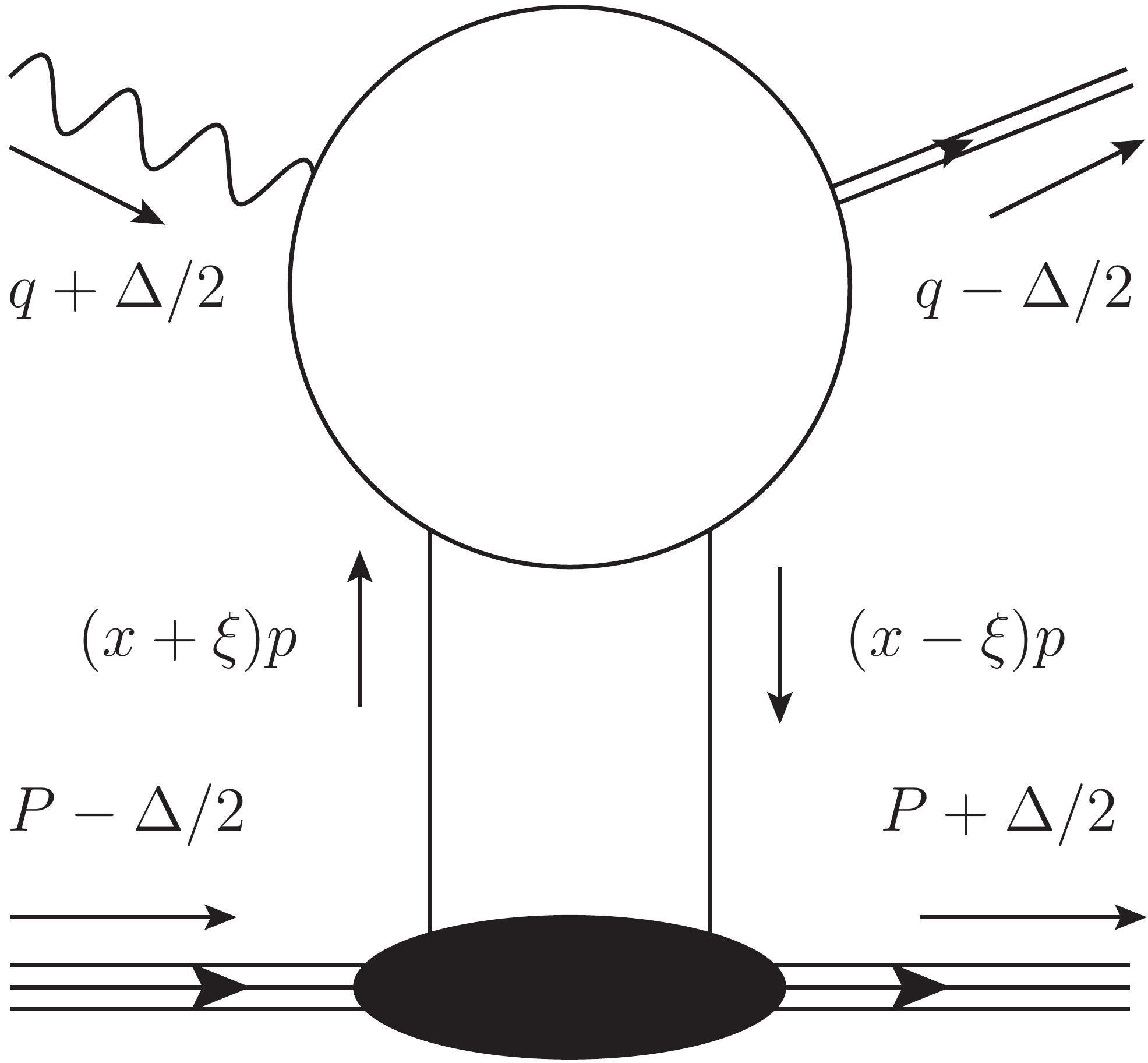}

  \caption{Schematic representation of HVM electroproduction.}
  \label{Blob}
  \end{center}
\end{figure}
We describe the matrix element for exclusive HVM electroproduction as the fluctuation of a hard incoming photon with momentum $q^\mu + \Delta^\mu/2$ into a heavy $Q \bar{Q}$ pair, which then interacts with the proton (or nuclei) carrying momentum $P^\mu-\Delta^\mu/2$ via a two-parton colour singlet exchange mechanism, see Fig.~\ref{Blob}. The proton recoils slightly with momentum $P^\mu+\Delta^\mu/2$. The modelling of the open quark-antiquark recombination into the observed exclusive final state HVM with momentum $q^\mu - \Delta^\mu/2$ is made, as in~\cite{Ivanov:2004vd, stethesis}, via LO Non-Relativistic-QCD~(NRQCD). In this approach, the amplitude for the production of two on-shell heavy quarks is calculated and projected onto the outgoing HVM quarkonium state. The amplitude can be expanded in powers of $\alpha_s$ and the heavy quark relative velocity. Here, we compute the $\alpha_s$ corrections. The relativistic corrections have been studied elsewhere, see~\cite{Hoodbhoy:1996zg}. To LO in the NRQCD relative velocity expansion, the momenta of the quark and anti-quark are equal such that their sum equals the momentum of the HVM. 

Following the set-up in~\cite{Ji:1998xh}, the three independent momenta defined above are decomposed in terms of high energy light-like Sudakov basis vectors~$\left\{p,n,\Delta_{\perp} \right\},$ satisfying $p \cdot p = n \cdot n = 0$ and $p \cdot n = 1$. The mean of the incoming and outgoing proton momenta, $P^\mu$,  defines the collinear direction. In the Bjorken limit to leading-twist accuracy, i.e.\ neglecting the masses of the nuclei, the kinematics of the process simplify and can be expressed in terms of the virtuality of the photon, $Q^2$, the mass, $M$, of the HVM, and the skewedness parameter, $\xi$. Here, $P^\mu \approx p^\mu$ and $\Delta^\mu \approx -2 \xi p^\mu,$ where $2 \xi$ is the `kick' which the active quark or gluon receives along the {\it collinear} direction so that the $t$-channel momentum exchange, $t = \Delta^2 = 0$.
The probed partons (gluons or quarks) carry momenta $p_1 = (x+\xi)\ p$ and $p_2 = -(x-\xi)\ p$, the momentum fraction $x$ is integrated over in the convolution with the Generalised Parton Distribution functions (GPDs).  
Inserting the replacements
\begin{equation}
    p^\mu \rightarrow \hat{p}^\mu = x p^\mu,\,\,\,\,\,\,\,n^\mu \rightarrow \hat{n}^\mu =  n^\mu/x,\,\,\,\,\,\,\,\xi \rightarrow \hat{\xi} = \xi/x,
\end{equation}
allow us to reduce the number of dimensionless variables appearing in our description by one. Note that these transformed basis vectors still respect $\hat{p} \cdot \hat{p} = \hat{n} \cdot \hat{n} = 0$ and $\hat{p} \cdot \hat{n} = 1$ and so all possible scalars formed from our momenta are unaffected by this change.

\begin{figure}[h!]
  \begin{center}
  \includegraphics[width=0.3\textwidth]{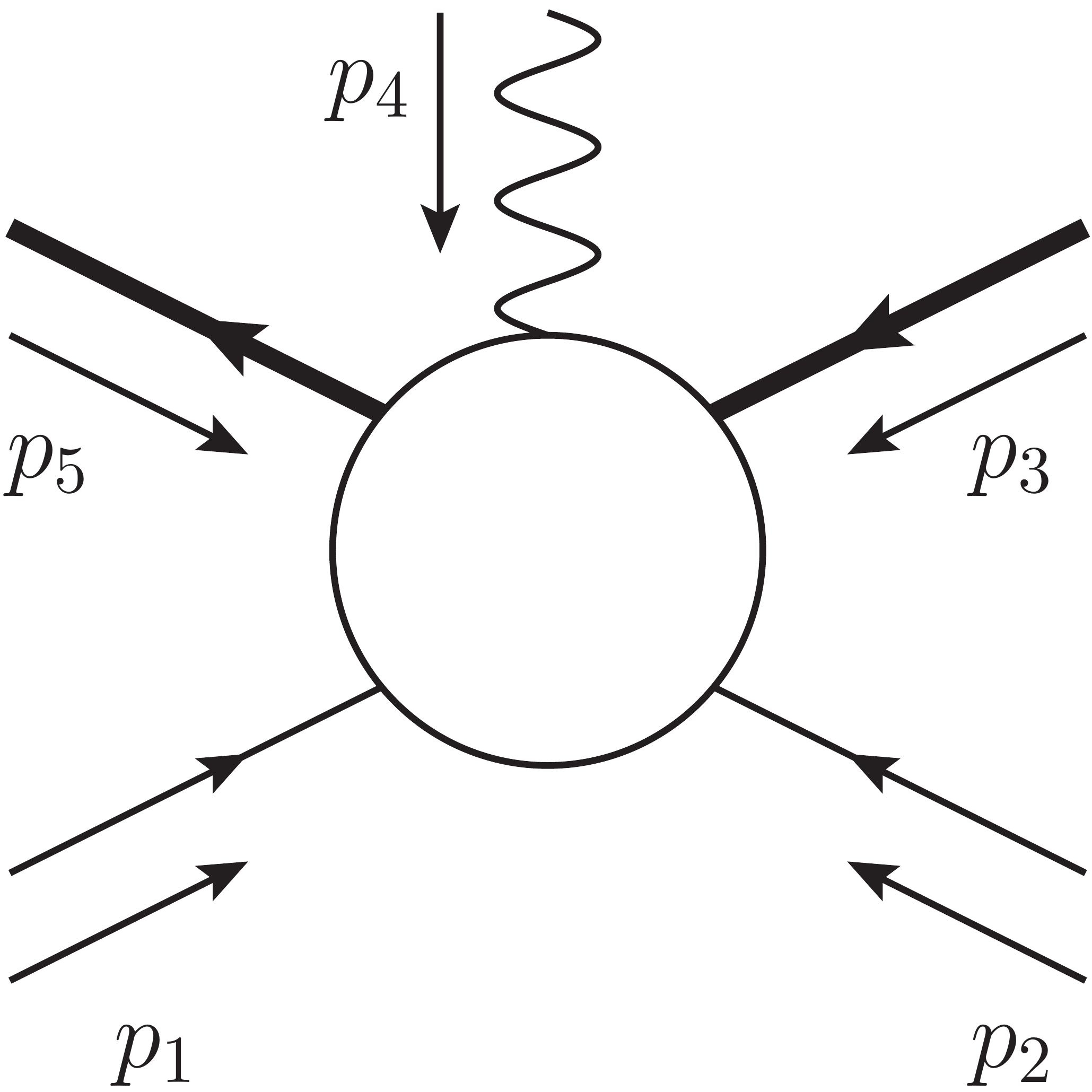}
  \qquad \qquad
  \includegraphics[width=0.3\textwidth]{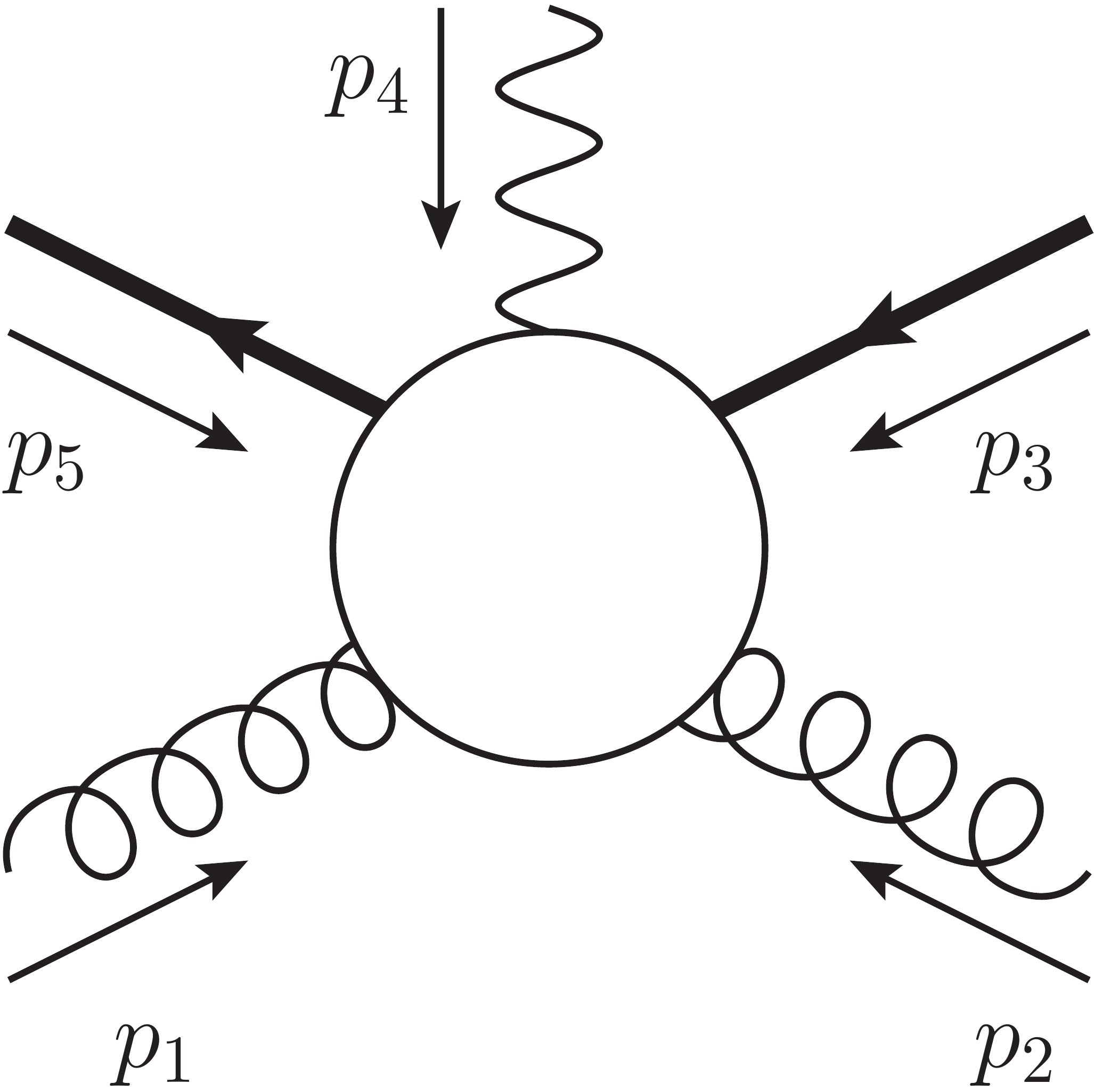}
  \caption{The kinematics of quark and gluon initiated processes. Massive quark lines are indicated in bold.}
  \label{HVMKin1}
  \end{center}
\end{figure}
Therefore, to leading order in the relative heavy quark velocity and in the Bjorken limit, the Sudakov decomposed momenta of Figure~\ref{HVMKin1} are\footnote{Henceforth, for notational simplicity, we will suppress the hat notation on the transformed Sudakov vectors $p$ and $n$.}
\begin{align}
p_1^\mu &= (1+r_1/r_3) p^\mu, &
p_2^\mu &=  -(1-r_1/r_3) p^\mu, \\
p_3^\mu = p_5^\mu &= \frac{(r_2-r_1)}{2r_3} p^\mu + \frac{r_3}{2} n^\mu, &
p_4^\mu &= -\frac{(r_1+r_2)}{r_3} p^\mu - r_3n^\mu,
\end{align}
where $$r_{1,2} = \frac{Q^2 \mp 4m^2}{4}\,\,\,\,\text{and}\,\,\,\,r_3 = \frac{Q^2 - 4m^2}{4 \hat{\xi}}.$$
Our convention is that all momenta are incoming. Moreover, in accordance with the leading term in the NRQCD expansion, we make the approximation that the on-shell pole mass of the heavy quarks is $m = M/2$.

At leading order only the gluon induced process contributes;
\begin{align}
\gamma^*(p_4) + g(p_1) \rightarrow Q(-p_5) + \bar{Q}(-p_3) + g(-p_2).
\end{align}

At NLO the $Q\bar{Q}$ pair may scatter from a light quark via a gluon exchange and so a new partonic channel opens. We compute in addition the quark induced process:
\begin{align}
&\gamma^*(p_4) + q(p_1) \rightarrow Q(-p_5) + \bar{Q}(-p_3) + q(-p_2), &(q=u,\bar{u},d,\bar{d},s,\bar{s}).
\end{align}

\subsection{HVM and GPD spin projections}

The S-wave, spin-triplet projection may be written to leading order in the heavy quark relative velocity as \cite{Petrelli:1997ge,Bodwin:2013zu,Braaten:2002fi}
\begin{equation}
v^i_\alpha(-p_3) \bar{u}^j_\beta(-p_5) \rightarrow 
\frac{-\delta^{ij}}{4 N_c}\frac{1}{4 m^2} 
\left( \frac{\langle O_1 \rangle_V}{m} \right)^{\frac{1}{2}} 
\left[ (- \slashed{p}_3 - m) \slashed{\epsilon}_S^* (-\slashed{K} + M) (-\slashed{p}_5 +m) \right]_{\alpha \beta}, \label{eq:HVMSpinProj}
\end{equation}
where, in the non-relativistic limit, we take the vector meson momenta $ K = 2 p_3 = 2 p_5$ and mass $M=2m$. Here $\bar{u}_{\beta}^j$ is the outgoing heavy quark spinor and $v_{\alpha}^i$ is the outgoing heavy anti-quark spinor. The indices $i$ and $j$ label the colour of the heavy quarks whilst $\alpha$ and $\beta$ label their spin. $\langle O_1 \rangle_V$ represents the non-perturbative NRQCD matrix element. The vector $\epsilon_S$ describes the polarisation of the HVM; it satisfies $\epsilon_S \cdot \epsilon_S^* = -1$ and $K \cdot \epsilon_S^* = 0$. Here relative to~\cite{Braaten:2002fi} we have an overall minus sign. It multiplies the overall amplitude and so has no effect on the cross-section. 

The quark GPD contraction is implemented as a spin projection of the on-shell quark scattering matrix. We replace the spinors of the quark and anti-quark connecting to the quark GPD, $F^q(x,\xi)$, by
\begin{equation}
u^i_\alpha(p_1) \bar{u}^j_\beta(-p_2) \rightarrow \frac{\delta^{ij}}{N_c}\frac{1}{2} F^q(x,\xi) \slashed{p}_{\alpha \beta}. \label{eq:GPDqSpinProjection}
\end{equation}
The factor $\slashed{p}_{\alpha \beta}$ results in a trace over the spin line of the quarks connecting to the GPD at the amplitude level. This can be understood by considering the numerator of a quark diagram. Representing the product of quark propagator numerators as $\left[ A \right]$ and applying the spin projection we obtain
\begin{equation}
\bar{u}_\beta(-p_2) \left[ A \right]_{\beta \alpha} u_\alpha(p_1) \rightarrow 
\left[ A \right]_{\beta \alpha} \slashed{p}_{\alpha \beta} =
\text{Tr}\left[ A \slashed{p} \right].
\end{equation}

Similarly, for the gluon induced partonic channel, the contraction with the gluon GPD, $F^g(x,\xi)$, is implemented as a projection of the on-shell gluon scattering matrix. 
Throughout we will use dimensional regularisation in $d = 4- 2 \epsilon$ space-time dimensions and make the replacement
\begin{equation}
\epsilon_1^{\mu} {\epsilon^*_2}^{\nu} \rightarrow - \frac{\delta_{ab}}{(N_c^2-1)} \frac{1}{2} \frac{1}{(d-2)} \frac{F^g(x,\xi)}{x_+ x_-} g_\perp^{\mu \nu}, \label{eq:GPDgPolProjection}
\end{equation}
where $\epsilon_1^{\mu}$ and ${\epsilon^*_2}^{\nu}$ are the polarisations of the incoming and outgoing gluons respectively.
Here, $x_+ = x+\xi - \mathrm{i} \delta$ and $x_- = x-\xi + \mathrm{i} \delta$. The correct $\mathrm{i} \delta$ prescription for the poles has been discussed extensively in the literature, see for example \cite{Braun:2002wu}. The prescription given here is valid for both DVCS and HVM production \cite{Pire:2011st,Ivanov:2004vd}. The indices $a$, $b$ are gluon colour indices in the adjoint representation. The factor $1/(N_c^2-1)$ averages over the gluon colours. The $1/(d-2)$ factor is our analytic continuation of the number of transverse polarisations of a gluon in $d$ dimensions. It appears due to the average over the gluon polarisations. The factor of $1/2$ is required to prevent double counting when both $s$ and $u$ channel gluon diagrams are computed (as done here) and the momentum fraction $x$ is integrated over from $-1$ to $1$ (see later). The object $g_\perp^{\mu \nu} \equiv g^{\mu \nu} - p^\mu n^\nu - p^\nu n^\mu$ carrying the gluons' Lorentz indices is the perpendicular metric tensor which projects onto the plane perpendicular to $p$ and $n$.

\subsection{Lorentz-invariant tensor decomposition}
We consider only the vector part of the amplitude at leading twist and at $t=0$. Higher-twist terms are formally suppressed and axial-vector contributions are neglected, as they are not needed here with an unpolarised nucleon in the initial state.
As shown in Section~\ref{setup}, in the Bjorken limit at leading twist, all of our external kinematics can be expressed in the Sudakov basis $\left\{p,n\right\}$ with $\Delta_{\perp} = 0.$ We decompose the part of the amplitude insensitive to the helicities of the incoming partons in the nucleon target in terms of the available Lorentz structure in this basis. Explicitly, we factor off the polarisation vectors for the incoming photon and outgoing HVM and work with the amputated amplitude $\mathcal T^{(\mu \nu)}$.\footnote{The round brackets $(\mu \nu)$ denote the vector part of the amplitude, $\mathcal T$, which is all that is needed in the description of an unpolarised measurement.} It follows that 
\begin{equation}
    \mathcal T^{(\mu \nu)} = A g^{\mu \nu} + B p^{\mu} n^{\nu} + C n^{\mu} p^{\nu} + D p^{\mu} p^{\nu} + E n^{\mu} n^{\nu}, 
\end{equation}
where $A,B,C,D$ and $E$ are the arbitrary coefficients of the decomposition. Imposing local current conservation at the photon vertex, $p_{4,\mu}  \mathcal T^{(\mu \nu)} = 0$, together with the identity $K_{\nu}  \mathcal T^{(\mu \nu)} = 0$, constrains the coefficients. The former is the familiar Ward-identity while the latter is, strictly speaking, not but holds at the Feynman diagram level and is true due to our choice of HVM spin projection. With these equations, we obtain the decomposition
\begin{align}
     \mathcal T^{(\mu \nu)} &= -g^{\mu \nu}_{\perp} T_{\perp} + \left( \frac{p_4 \cdot p}{p_4 \cdot n} n^{\mu} - p^{\mu} \right) \left( \frac{p_3 \cdot p}{p_3 \cdot n} n^{\nu} - p^{\nu} \right) \frac{\tilde{T}_L}{4} \label{tensordecomp1} \\  &= -g^{\mu \nu}_{\perp} T_{\perp} + \ell^{\mu \nu} T_L, \label{tensordecomp2}
\end{align}
where \begin{equation} 
\ell^{\mu \nu} = \frac{\mathcal N}{4} \left( \alpha n^{\mu} - p^{\mu} \right) \left(\beta n^{\nu} - p^{\nu} \right)\,\,\,\,\,\,\,\,\text{and}\,\,\,\,\,\,\,\,T_L =~ \tilde{T}_L/ \mathcal N,
\label{eq:N}
\end{equation} with \begin{equation} \alpha \equiv \frac{p_4 \cdot p}{p_4 \cdot n}\,\,\,\,\,\,\,\,\text{and}\,\,\,\,\,\,\,\,\beta \equiv \frac{p_3 \cdot p}{p_3 \cdot n}.\end{equation}

An explicit exposition of the available Lorentz structure has therefore produced a manifest decoupling of the system into two overarching degrees of freedom, parametrised by $T_{\perp}$ and $T_L$. Contractions of~(\ref{tensordecomp2}) with explicit realisations of the physical transverse and longitudinal polarisation vectors of the photon $\varepsilon^{\gamma}_{\mu}$ and HVM $\varepsilon^{V*}_{\nu}$ pick out one of the two scalar coefficients in each case. This may be seen as follows. The transverse polarisation vectors have only a transverse component in their Sudakov-basis decomposition. By construction, their contraction with $p^\mu, p^\nu, n^\mu, n^\nu$ in $\ell^{\mu \nu}$ vanishes and the only contribution is due to $g_{\perp}^{\mu \nu}:$
\begin{equation}
    \varepsilon^{\gamma}_{\pm,\mu} \varepsilon^{V*}_{\pm, \nu}\, \mathcal{T}^{(\mu \nu)} = -\varepsilon^{\gamma}_{\pm,\mu} \varepsilon^{V*}_{\pm, \nu} \, g_{\perp}^{\mu \nu}\, T_{\perp} = -\varepsilon^{\gamma}_{\pm} \cdot \varepsilon^{V*}_{\pm}\, T_{\perp} = T_{\perp},
\end{equation}
where $\varepsilon^{\gamma}_{\pm,\mu}$ and $\varepsilon^{V*}_{\pm, \nu}$ are the transverse polarisation vectors for the photon and HVM in the helicity basis, respectively. The corresponding longitudinal polarisation vectors are 
\begin{align}
    \varepsilon^{\gamma, \mu}_{L} &= \frac{2Q\hat{\xi}}{M^2-Q^2} p^{\mu} + \frac{(Q^2-M^2)}{4 Q \hat{\xi}} n^{\mu}, \\
     \varepsilon^{V,\nu *}_{L} &= \frac{2M\hat{\xi}}{M^2-Q^2} p^{\nu} + \frac{(Q^2-M^2)}{4 M \hat{\xi}} n^{\nu},
\end{align}
satisfying $\varepsilon^{\gamma}_{L} \cdot \varepsilon^{\gamma}_{L} = -1$ and $p_4 \cdot  \varepsilon^{\gamma}_{L} = 0$, with similar relations for the HVM.  Then,
\begin{equation}
    \varepsilon^{\gamma}_{L,\mu} \varepsilon^{V*}_{L,\nu}\, \mathcal{T}^{(\mu \nu)} = \varepsilon^{\gamma}_{L,\mu} \varepsilon^{V*}_{L, \nu} \, \ell^{\mu \nu}\, T_{L} = T_L,
\end{equation}
where there is now no contribution from $g_{\perp}^{\mu \nu}$ and where we have fixed $\mathcal N$ in~(\ref{eq:N}) such that the contraction of $\ell^{\mu \nu}$ with the physical longitudinal polarisation vectors is unity. 

The decomposition~(\ref{tensordecomp1},\ref{tensordecomp2}) is therefore readily identifiable as a separation into transverse and longitudinal components, with $T_{\perp}$ and $T_L$ having the physical interpretation now as the process's transverse and longitudinal form factors, respectively.  
With this choice\footnote{This degree of freedom is evident in~(\ref{eq:N}), where the introduction of $\mathcal{N}$ allows for a re-shuffling of terms between $\ell^{\mu \nu}$ and $T_L$.}, the transverse and longitudinal helicity amplitudes, 
$A^{\pm \pm}=\varepsilon^{\gamma}_{\pm,\mu} \varepsilon^{V*}_{\pm,\nu} \mathcal{T}^{(\mu \nu)}$ and $A^{00} = \varepsilon^{\gamma}_{L,\mu} \varepsilon^{V*}_{L,\nu}\, \mathcal{T}^{(\mu \nu)}$, are equal to $T_{\perp}$ and $T_L$, respectively.  Note also that the introduction of the $\mathcal N$ factor into $\ell^{\mu \nu}$ in~(\ref{eq:N}) allows for both two-tensors in multiplication with the scalar coefficients $T_{\perp}$ and $T_L$ to have mass dimension zero. In this way, one may extract these coefficients in turn through suitable projections onto the $\mathcal T^{(\mu \nu)}$ structure.

We remark that~(\ref{tensordecomp1}) coincides with the leading-twist tensor decomposition found in Generalised-Deeply-Virtual-Compton-Scattering~(GDVCS), see e.g.~\cite{Belitsky:2012ch}, upon neglecting the axial-vector and helicity flip contributions. This is to be expected as the only distinction in the kinematical set-up is the final state production of a heavy photon instead of a heavy vector meson that we have here, however this remains indifferent to the construction of the underlying tensorial structure and the applicability of our Ward and Ward-like identities.

The vector part of the amplitude may be written, using collinear factorisation, as 
\begin{align}
\begin{split}
\mathcal{T}^{(\mu \nu)} =& 
-g^{\mu \nu}_\perp \int_{-1}^1 \frac{ \text{d} x}{x} \left[ \sum_q F^q(x,\xi) C_{\perp,q}\left(\frac{\xi}{x}, Q^2 \right) + C_{\perp,g}   \left(\frac{\xi}{x}, Q^2 \right) \frac{F^g(x,\xi)}{x} \right]\\
&+\ell^{\mu \nu} \int_{-1}^1 \frac{\text{d} x}{x} \left[ \sum_q F^q(x,\xi) C_{L,q}\left(\frac{\xi}{x}, Q^2 \right) + C_{L,g}  \left(\frac{\xi}{x}, Q^2 \right) \frac{F^g(x,\xi)}{x} \right] \\
&+ \ldots. 
\end{split} \label{eq:HVMPartonFac}
\end{align}
where the ellipses represent contributions that only appear beyond the leading twist term and outwith the chiral-even theory with $t=0$, but which would appear in e.g. polarised scattering or if the nucleon mass would not be neglected. The renormalised quark and gluon GPDs are denoted $F^q$ and $F^g$ respectively. $C_{\perp,q}$ and $C_{\perp,g}$ are the renormalised quark and gluon vector transverse coefficient functions while $C_{L,q}$ and $C_{L,g}$ are the renormalised quark and gluon vector longitudinal coefficient functions. The dependence of the GPDs $F^q$ and $F^g$ on the factorisation scale $\mu_F$ and on $t$ has been suppressed. The dependence of the renormalised coefficient functions $C_{\perp,q}, C_{\perp,g}, C_{L,q}$ and  $C_{L,g}$ on $m$, $\mu_F$ and the renormalisation scale $\mu_R$ has been suppressed, too. We recall that the Lorentz indices $\mu$ and $\nu$ are those of the incoming photon and outgoing HVM respectively. Measurements of HVM production from unpolarised targets probe only the charge conjugation even quark GPD;  we may therefore replace $\sum_q F^q$ with the singlet quark GPD $F^S/2$. There is also an additional photon helicity flip term which we neglect. 

\section{Overview of the calculation}
\begin{figure}[h!]
 \label{diags}
  \begin{center}
  \includegraphics[width=0.27\textwidth]{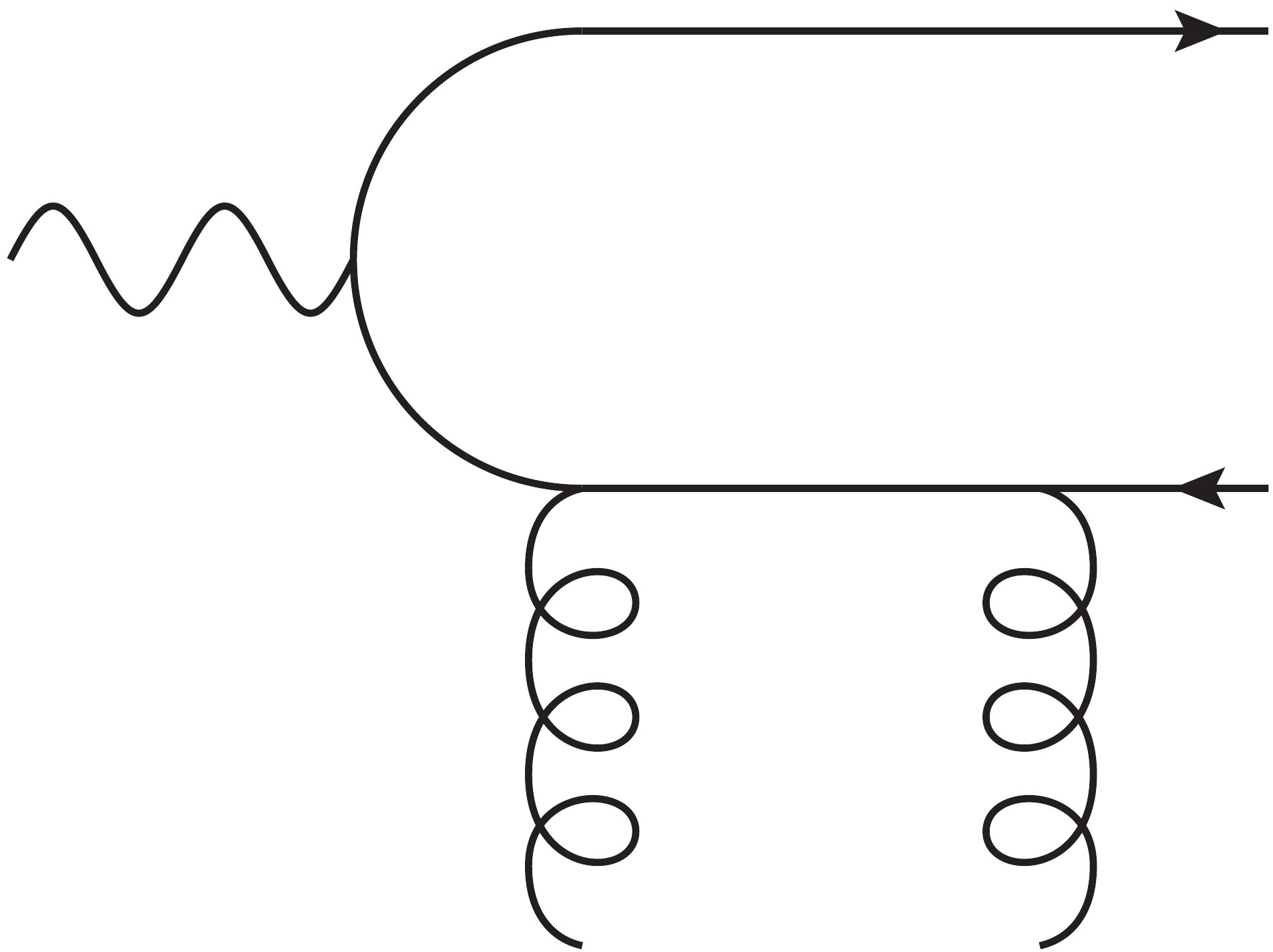}
  \qquad
  \includegraphics[width=0.27\textwidth]{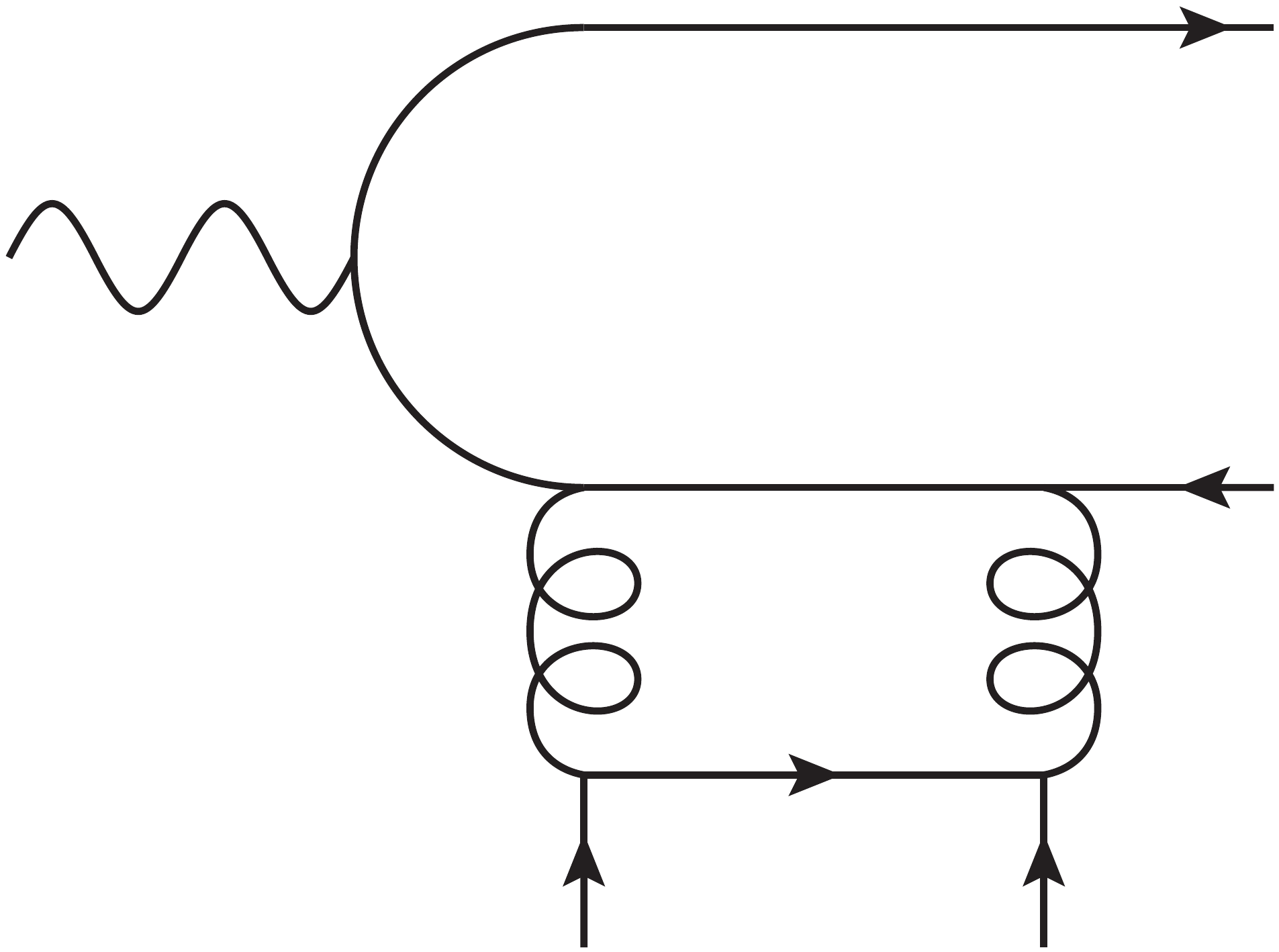}
   \qquad
  \includegraphics[width=0.27\textwidth]{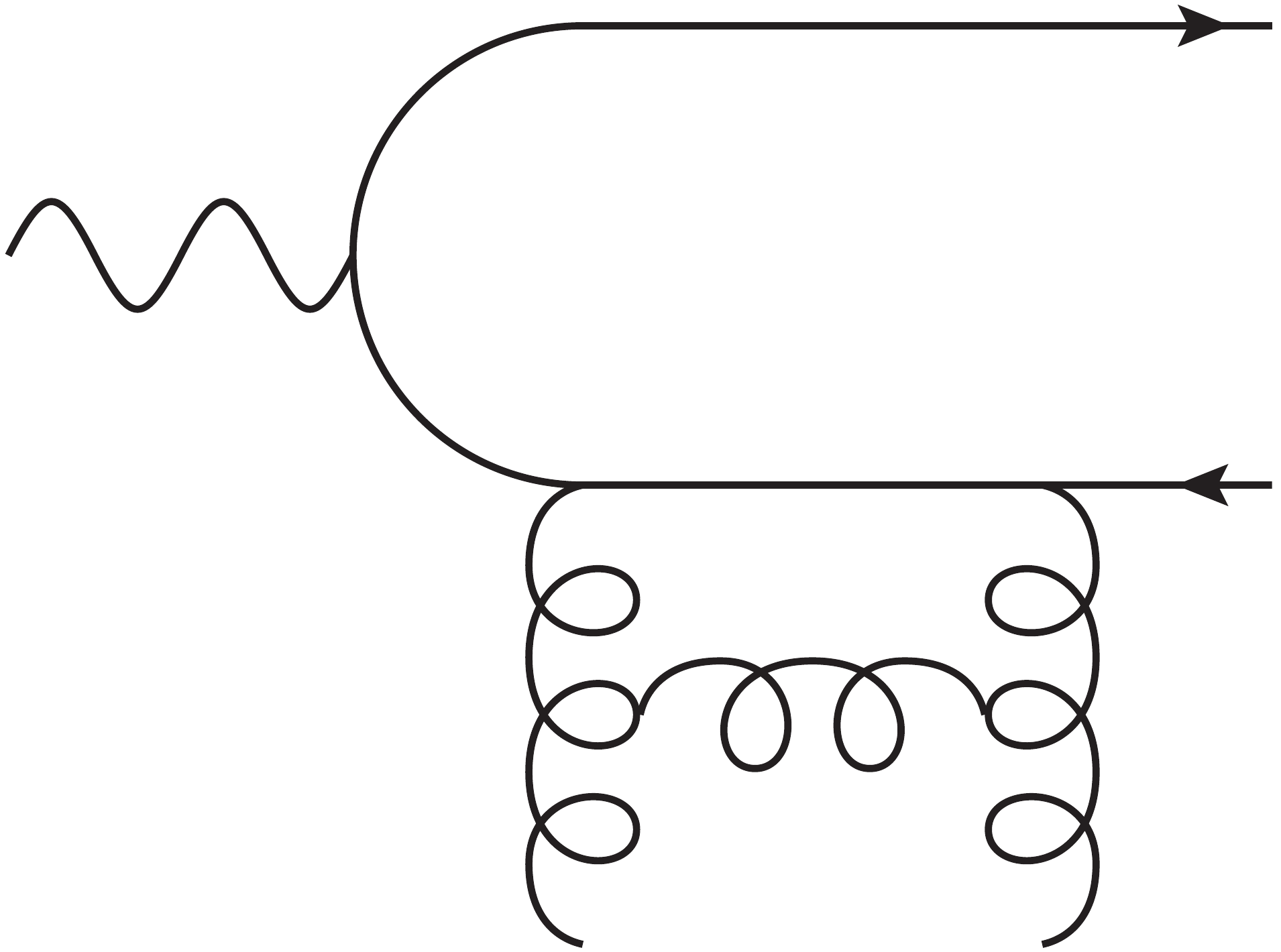}
  \caption{From left to right: LO gluon, NLO quark and NLO gluon initiated subprocess Feynman diagrams.}
  \end{center}
\end{figure}

We generate all LO and NLO Feynman diagrams using $\texttt{QGRAF}$~\cite{Nogueira:1991ex} and select those diagrams that are compatible with our external colour and kinematical constraints. Example Feynman diagrams for the LO and NLO quark and gluon initiated subprocesses are shown in Fig~\ref{diags}.   Each selected diagram is converted into an expression through insertion of Feynman rules, derived in an arbitrary linear covariant gauge. The appropriate GPD quark or gluon projector is applied to each diagram, together with the HVM spin projection, and the resulting $d$-dimensional Dirac traces are computed and handled in $\texttt{FORM4.2}$~\cite{Ruijl:2017dtg}. As our amplitude is written with two free Lorentz indices, $\mu$ and $\nu$, we need at most a basis decomposition for the rank-two tensor $k^{\mu} k^{\nu},$ where $k$ is the loop momentum. 

Due to the external colour and kinematical constraints, the integral structures obtained contain in general linearly {\it dependent} propagators. They therefore cannot be reduced directly using standard integral reduction tools. We express them first as structures containing only linearly {\it independent} propagators by applying a partial fractioning routine in line with the Leinartas algorithm~\cite{Lei78}. In this way, we eliminate the linear dependence amongst the propagators and proceed with the integral reduction via \texttt{REDUZE2}~\cite{vonManteuffel:2012np} which encodes Laporta's integration by parts algorithm~\cite{Laporta:2001dd}.

All results presented below are given in the ERBL region, $| \hat{\xi}| > 1$ (i.e. $|x| < \xi$), where we expect an absence of imaginary parts. The results in the physical region, $|\hat{\xi}| < 1$ (i.e. $|x| > \xi$), may be obtained by restoring the correct analytic continuation. Explicitly, this is 
\begin{equation}
\hat{\xi} \rightarrow \hat{\xi} - \text{sgn}(\hat{\xi})\, \mathrm{i}\delta\,\,\,\,\text{with}\,\,\,\,\delta \rightarrow 0^+.
\label{analytic}
\end{equation}
The coefficient functions are expanded in the bare strong coupling $a_0 = \alpha_0/(4 \pi)$ as,
\begin{align}
\hat{C}_{i,j} = a_0\ \hat{C}^{(0)}_{i,j} + a_0^2\ \hat{C}^{(1)}_{i,j} + \mathcal{O}(a_0^3), \label{eq:asexp}
\end{align}
with $i=\perp,L$ and $j=q,g$.

\section{LO results}
At leading-order~(LO) in $\alpha_s$, there is only a gluon initiated subprocess. There is no contribution from an initial state light quark as it only couples to the outgoing heavy quark system via a loop-induced gluon insertion at next-to-leading-order~(NLO). A single gluon exchange is $\propto f^{abc}$, which vanishes due to our GPD spin projection $\propto \delta^{ab}$ and antisymmetry of the SU(3) structure constants. The tree level contribution is therefore dominated by a gluon-gluon (pomeron-like) exchange: 
\begin{align}
\gamma^*(p_4) + g(p_1) \rightarrow Q(-p_5) + \bar{Q}(-p_3) + g(-p_2).
\end{align}

With the expansion in the bare strong coupling defined in \eqref{eq:asexp}, the bare (denoted by hat script) transverse and longitudinal coefficient functions at LO are 
\begin{equation}
\hat{C}_{\perp, q}^{(0)} = \hat{C}_{L, q}^{(0)} = 0, \label{tree3} \end{equation}
while
\begin{equation}
\hat{C}_{\perp, g}^{(0)} = -\sqrt{w^2-1} \, \hat{C}_{L, g}^{(0)} = \frac{2(w^2-1)}{w^2} \frac{T_F}{N_c} B_0.
\label{tree1} \end{equation}
Here, \begin{equation}B_0 = 16 \pi^2 g_e e_q  \left(\frac{ \langle O_1 \rangle_V}{m^3} \right)^{\frac{1}{2}}\frac{x^2}{x_+ x_-},\end{equation} where $g_e$ ($g_s$) is the electromagnetic (strong) coupling, $e_q$ is the photon-quark charge and $\langle O_1 \rangle_V$ is the NRQCD matrix element. The colour factor $T_F = 1/2$,  and
\begin{equation}
    w = \sqrt{1-\frac{4m^2}{Q^2}}.
\end{equation}
 
Eqns.~(\ref{tree3}-\ref{tree1}) are true up to and including $\mathcal O(\epsilon)$. There is a cancellation of $\epsilon$ terms between those appearing in the $1/(d-2)$ factor of eqn.~(\ref{eq:GPDgPolProjection}) and those in the numerators of the Feynman diagrams. This observation is important as the tree level results enter into the renormalisation of the NLO result. We will therefore not generate finite surplus contributions of the form $\epsilon/\epsilon$ in the computation of the NLO counterterms, see Section~\ref{renorm}.
 
The $Q^2 \rightarrow 0$ (i.e. $w \rightarrow \infty$ ) limit of these results coincide with the photoproduction result, \cite{Ivanov:2004vd, stethesis}, where only the amplitude to produce a transversely polarised HVM is non-vanishing.\footnote{Away from $t=0$, the quantum numbers of the photon and the HVM need not be the same so there is an (albeit suppressed) amplitude to produce a longitudinally polarised HVM from a transversely polarised photon.} Note that the limiting point $w \rightarrow 0$ is not kinematically attainable. The production of a time-like vector meson (with invariant mass squared, $M^2 = 4m^2 > 0$)  initiated by a space-like photon, $Q^2<0$, does not, as expected within the analytic structure of the $S$-matrix, produce a pole at LO.

\section{NLO results}

At NLO, there are quark and gluon initiated subprocesses. We compute
\begin{align}
&\gamma^*(p_4) + g(p_1) \rightarrow Q(-p_5) + \bar{Q}(-p_3) + g(-p_2), \\
&\gamma^*(p_4) + q(p_1) \rightarrow Q(-p_5) + \bar{Q}(-p_3) + q(-p_2),\,\,\,\,\,\,\,\,\,(q=u,\bar{u},d,\bar{d},s,\bar{s}),
\end{align}
and express our bare quark and gluon NLO coefficient functions in terms of sets of coefficients $c_i$ and a universal basis set $\left\{f_i,\, i=1,\dots,12 \right\}$ of logarithms and dilogarithms that arise from the Feynman loop integrations.
They are 
\begin{equation}
    \begin{split}
        f_1 &= 1,\,\,\,\,f_2 = \ln \left( \frac{w-1}{w+1} \right),\,\,\,\,f_3 = f_2^2,\,\,\,\,f_4 = \ln \left( \frac{2w^2}{w^2-1} \right), \\
        f_5 &=f_4^2,\,\,\,
        f_6 = \ln \left( \frac{2vw^2}{vw^2-1} \right),\,\,\,\, f_7 = 2 f_4 f_6 - f_6^2, \\
        f_8 &= \text{Li}_2 \left( \frac{1-v}{v(w^2-1)}\right),\,\,\,\,f_9 =\text{Li}_2 \left( \frac{1+w^2}{1-w^2}\right), \\
        f_{10} &= \text{Li}_2 \left( -\frac{1}{vw^2} \right) + \text{Li}_2 \left( \frac{1+v}{1-vw} \right) + \text{Li}_2 \left( \frac{1+v}{1+vw} \right) + \text{Li}_2 \left( \frac{v-1}{v^2w^2-1} \right),\\
        f_{11} &= \sqrt{-1+\frac{2(1+v)}{1+vw^2}} \, \ln \left(\frac{1+vw^2}{v(w^2-1)}\left(\frac{1+v}{1+vw^2} + \sqrt{-1+\frac{2(1+v)}{1+vw^2}} \right) \right), \\
        f_{12} &= f_{11}^2.
    \end{split}
    \label{fiset}
\end{equation}
Here, $v = \hat{\xi}/w^2$ and so, with this choice of variables, the analytic continuation to the DGLAP regime, as specified in eqn.~(\ref{analytic}), is restored via 
\begin{equation}
v \rightarrow v - \text{sgn}(v)\mathrm{i} \delta\,\,\,\text{with}\,\,\,\delta \rightarrow 0^+\,\,\,\text{and where}\,\,\,\text{sgn}(v) = \text{sgn}(\hat{\xi}).
\end{equation}

\subsection{Quark subprocess}
The quark $q$ subprocess amplitude, $A_q$, to NLO can be written in the form
\begin{equation}
A_q = \sum_{a=1}^6 A_{a,q}\,\,\,\,\,\,\,\,\,\,\text{and where}\,\,\,\,\,\,\,\,\,\
A_{a,q} \sim \int \frac{\text{d}^d k}{(2 \pi)^d}\, \text{Tr} \left( \chi_1 \Gamma^{(1)}_a \right) \, \text{Tr} \left( \chi_2 \Gamma^{(2)}_a \right).
\end{equation}
 Here, $\chi_{1,2}$ are the HVM and GPD projectors and $\Gamma^{(1,2)}_a$ encode the Dirac structure for each spin line in diagram $a$. The appearance of a trace at amplitude level is particularly noteworthy and is a reflection of our external colour and kinematical constraints.

Each of the four quark diagrams where the external photon couples to the open heavy quark or antiquark lines may be decomposed into a tadpole and two bubble master integrals that emerge from the Laporta algorithm. When the photon instead attaches to the heavy quark mass propagator, there are in addition two triangle master integrals in the diagram decomposition. In this latter case, we have decomposed a pentagon integral, containing linearly dependent propagators, into a sum of triangles, bubbles and the tadpole integral which, by construction, contain only linearly independent propagators.  

We normalise our quark NLO coefficient function with the factor
\begin{equation}
    A_1 = (4 \pi)^4 g_e e_q \left( \frac{ \langle O_1 \rangle_V}{m^3} \right)^{1/2} C_{\epsilon} \left(\frac{\mu_0^2}{m^2} \right)^\epsilon,
    \label{norm1}
\end{equation}
where 
\begin{equation}
    C_{\epsilon} = \frac{\pi^{d/2}}{(2 \pi)^d} \frac{\Gamma^2(d/2-1) \Gamma(3-d/2)}{\Gamma(d-3)} = \frac{1}{16 \pi^2} \biggl( 1- \epsilon (\gamma_E - \ln(4 \pi) ) \biggr) + \mathcal O(\epsilon^2)
    \label{Ceps1}
\end{equation}
arises from our loop integrals and the scale $\mu_0$ is a mass parameter introduced in dimensional regularisation to maintain a dimensionless bare coupling. Here, $\gamma_E$ is the Euler-Mascheroni constant.
The bare transverse quark NLO coefficient function can be written in the form 
\begin{equation}
    \hat{C}_{\perp,q}^{(1)} = \frac{T_F^2}{N_c^2}A_1 \left( \sum_{i=1}^{12} c_{\perp,i} f_i + \left\{ v \rightarrow -v \right\} \right),
\end{equation}
where the $f_i$ are given in eqn.~(\ref{fiset}) and the $c_{\perp,i}$ are
{\allowdisplaybreaks \small
\begin{equation}
\begin{split}
    c_{\perp,1} = &\pi ^2 \left(\frac{-8 v^2+16 v-8}{-1+v w^2}+\frac{8 v^2+16 v+8}{1+v
   w^2}+\frac{16}{w^4}-\frac{32}{w^2}\right), \\
   c_{\perp,2} = &\frac{32}{w^3}-\frac{32}{w},\,\,\,\, c_{\perp,3} = \frac{-12 v^2+24 v-12}{-1+v w^2}+\frac{12 v^2+24 v+12}{1+v w^2}+\frac{24}{w^4}-\frac{48}{w^2}, \\
   c_{\perp,4} = &-\frac{32}{(v+1) \left(w^2+1\right)}+\frac{16}{(v+1) w^2}+\frac{32}{(v-1)
   \left(w^2+1\right)}-\frac{16}{(v-1) w^2}+\frac{32}{w^2}-32, \\
    c_{\perp,5} = &0,\,\,\,\, c_{\perp,6} =\frac{1}{\epsilon
   } \left( \frac{64}{v w^4}+\frac{64 v+64}{1+v w^2}-\frac{64}{v w^2}-\frac{64}{w^2} \right) +\frac{-32 v-64}{1+v w^2}-\frac{64}{(v-1) \left(1+v w^2\right)} \\& +\frac{32}{(v-1) w^2}+32,\,\,\,\,c_{\perp,7} = -\frac{64}{v w^4}+\frac{-64 v-64}{1+v w^2}+\frac{64}{v w^2}+\frac{64}{w^2},\,\,\,\,c_{\perp,8} = -c_{\perp,7},\\
   c_{\perp,9} = &\frac{-24 v^2+16 v+8}{-1+v w^2}+\frac{24 v^2+16 v-8}{1+v w^2}+\frac{48}{w^4}-\frac{32}{w^2},\\c_{\perp,10} = &\frac{-48 v^2+96 v-48}{-1+v w^2}+\frac{-48 v^2-96 v-48}{1+v w^2}-\frac{64}{v w^4}+\frac{64
   v}{w^2}+\frac{64}{v w^2}+32 v,\\c_{\perp,11} &=0,\,\,\,\,c_{\perp,12} =0.
    \end{split}
    \label{quarktrans}
\end{equation}}
Similarly, the bare longitudinal quark NLO coefficient function can be expressed as 
\begin{equation}
    \hat{C}_{L,q}^{(1)} = \frac{T_F^2}{N_c^2}A_1 \sqrt{w^2-1} \left( \sum_{i=1}^{12} c_{L,i} f_i + \left\{ v \rightarrow -v \right\} \right),
\end{equation}
where the $f_i$ are given in eqn.~(\ref{fiset}) and the $c_{L,i}$ are
{\allowdisplaybreaks \small
\begin{equation}
\begin{split}
    c_{L,1} &= 0,\,\,\,\,c_{L,2} = \frac{32}{w}-\frac{64}{w^3},\,\,\,\,c_{L,3} = 0, \\ c_{L,4} = & \frac{16}{(v+1) w^2}-\frac{16}{(v-1) w^2}-\frac{16}{v+1}+\frac{16}{v-1}-\frac{64}{w^2}+32, \\ c_{L,5} = &0,\,\,\,\, c_{L,6} = \frac{1}{\epsilon } \left( {\frac{64}{v w^4}+\frac{64 v}{1+v w^2}}-\frac{64}{w^2}  \right) + \frac{64 v}{1+v w^2} +\frac{32}{(v-1) w^2}-\frac{32}{v-1}-32,\\ c_{L,7} = &-\frac{64}{v w^4}-\frac{64 v}{1+v w^2}+\frac{64}{w^2},\,\,\,\,c_{L,8} = -c_{L,7},\,\,\,\,c_{L,9} = -\frac{32 v}{-1+v w^2}-\frac{32 v}{1+v w^2}+\frac{64}{w^2},\\ c_{L,10} = &-\frac{64}{v w^4}+\frac{64 v}{w^2}-32 v,\,\,\,\,c_{L,11} = 0,\,\,\,\,c_{L,12} = 0.
    \end{split}
    \label{quarklong}
\end{equation}}

The expressions for both the bare quark transverse and longitudinal coefficient functions are written in a manifestly symmetrised form, where $v \rightarrow -v$ corresponds to the physical $\hat{\xi} \rightarrow -\hat{\xi}$ symmetry. The expressions have been expanded in $\epsilon$, retaining the singular term in $1/\epsilon$ and the finite term, while neglecting $\mathcal O(\epsilon)$ terms which are not required at this order.

\subsection{Gluon subprocess}
We normalise our gluon NLO coefficient function with the factor
\begin{equation}
    B_1 = (4 \pi)^4 g_e e_q \left( \frac{ \langle O_1 \rangle_V}{m^3} \right)^{1/2}\frac{x^2}{x_+ x_-} C_{\epsilon} \left(\frac{\mu_0^2}{m^2} \right)^\epsilon,
\end{equation}
where $C_{\epsilon}$ is defined in eqn.~(\ref{Ceps1}).

The bare transverse gluon NLO coefficient function can be written in the form 
\begin{equation}
    \hat{C}_{\perp,g}^{(1)} =  \frac{B_1}{N_c (N_c^2-1)} \left( \sum_{i=1}^{12} c_{\perp,i} f_i + \left\{ v \rightarrow -v \right\} \right),
\end{equation}
where the $f_i$ are given in eqn.~(\ref{fiset}) and the $c_{\perp,i}$ are
{\allowdisplaybreaks \small
\begin{align*}
c_{\perp,1} = &\frac{C_F}{\epsilon} \left(-6 -\frac{12}{w^4} + \frac{6}{w^2} + \frac{-6+6v}{-1+vw^2} + \frac{6+6v}{1+vw^2} \right) +  C_F \biggl(-32 + \frac{2}{1+v} - \frac{2}{-1+v} \\&+ \pi^2 \left( -16 + \frac{16}{3w^4} - \frac{22-80v^2}{3w^2}\right) - \frac{16}{w^4} + \frac{40}{w^2} - \frac{2}{(1+v)w^2} + \frac{2}{(-1+v)w^2} + \frac{8}{1+w^2} 
 \\&+ \frac{\pi^2 \left(-\frac{22}{3} + \frac{85v}{3} - \frac{104 v^2}{3} + \frac{41 v^3}{3} \right) }{ (-1+vw^2)^2 } - \frac{2-2v}{-1+vw^2} + \frac{\pi^2 \left(-\frac{49}{3} + 32v - \frac{7 v^2}{3} - \frac{40 v^3}{3} \right) }{ -1+vw^2 } \\&+  \frac{\pi^2 \left(-\frac{22}{3} - \frac{85v}{3} - \frac{104 v^2}{3} - \frac{41 v^3}{3} \right) }{ (1+vw^2)^2 }  + \frac{2+2v}{1+vw^2} +  \frac{\pi^2 \left(\frac{49}{3} + 32v + \frac{7 v^2}{3} - \frac{40 v^3}{3} \right) }{ 1+vw^2 } \biggr) \\& + C_A \biggl( \pi^2 \left( \frac{22}{3} - \frac{4}{w^4} + \frac{8(1-v^2)}{w^2} \right) +  \frac{\pi^2 \left(\frac{10}{3} - \frac{37v}{3} + \frac{44 v^2}{3} - \frac{17 v^3}{3} \right) }{ (-1+vw^2)^2 } \\&+  \frac{\pi^2 \left(9 - \frac{58v}{3} + \frac{19 v^2}{3} + 4v^3 \right) }{ -1+vw^2 } + \frac{\pi^2 \left(\frac{10}{3}+ \frac{37v}{3} + \frac{44v^2}{3} + \frac{17v^3}{3}\right) }{ (1+vw^2)^2 } \\&+  \frac{\pi^2 \left(-9 - \frac{58v}{3} - \frac{19 v^2}{3} + 4v^3 \right) }{ 1+vw^2 } \biggr), \\
c_{\perp,2} =  &C_F \biggl(8 v^2 w^3-24 v^2 w+\frac{128
   v^2}{w}+\frac{-56 v^3 w + 52 v^2 w  -4 v w}{-1+v w^2}+\frac{-56 v^3 w-52 v^2 w-4 v w}{1+v
   w^2} \\&+\frac{40}{w^3}-\frac{40}{w}\biggr) + C_A \biggl(-\frac{64 v^2}{w}+\frac{32 v^3 w-52 v^2 w+20 v w}{-1+v w^2} +\frac{32 v^3 w+52 v^2
   w+20 v w}{1+v w^2}\\&-\frac{8}{w^3}+\frac{8}{w}\biggr), \\
c_{\perp,3} = &C_F \biggl(\frac{32 v^2}{w^2}+\frac{-16 v^3+32 v^2-24 v+8}{-1+v w^2}+\frac{-16 v^3-32 v^2-24
   v-8}{1+v w^2}+\frac{-3 v^3+8 v^2-7 v+2}{\left(-1+v w^2\right)^2} \\&+\frac{3 v^3+8 v^2+7 v+2}{\left(1+v
   w^2\right)^2}+\frac{10}{w^4}-\frac{12}{w^2}+14\biggr) + C_A \biggl(-\frac{8 v^2}{w^2}+\frac{4 v^3-6 v^2+2}{-1+v w^2}+\frac{4 v^3+6 v^2-2}{1+v
   w^2}\\&+\frac{3 v^3-8 v^2+7 v-2}{\left(-1+v w^2\right)^2}+\frac{-3 v^3-8 v^2-7 v-2}{\left(1+v
   w^2\right)^2}-\frac{6}{w^4}+\frac{12}{w^2}+2\biggr), \\
c_{\perp,4} = &C_F \biggl(8 v^2 w^4-24 v^2 w^2+\frac{72 v^2-44 v-16}{-1+v w^2}+\frac{8 v^2-52 v-56}{1+v
   w^2}+16 v^2-4 v w^4+20 v w^2\\&+\frac{4 w^2+\frac{4}{w^2}-8}{(v+1)^2}+\frac{8}{(v+1)
   \left(w^2+1\right)}-\frac{8}{(v+1) \left(v w^2-1\right)}+\frac{8}{\left(w^2+1\right) \left(v
   w^2+1\right)}-\frac{64 v}{w^2}\\&+\frac{-4 w^4-8 w^2-\frac{8}{w^2}+12}{v+1}-32 v+4 w^4+4
   w^2-\frac{72}{w^2+1}+\frac{16}{\left(w^2+1\right)^2}+\frac{56}{w^2}\biggr)\\&+C_A\biggl(\frac{-8 v^2+20 v-12}{-1+v w^2}+\frac{8 v^2+4 v-4}{1+v
   w^2}+\frac{8-\frac{8}{w^2}}{v+1}+\frac{16}{w^2+1}-\frac{8}{w^2}-8\biggr),\\
c_{\perp,5} =   &C_A \left(\frac{16}{v w^4}+\frac{16-16 v}{-1+v w^2}+\frac{16 v+16}{1+v w^2}-\frac{16}{v
   w^2}\right), \\
c_{\perp,6} = &\frac{C_A}{\epsilon} \left(-\frac{16}{v w^4}-\frac{32 v+32}{v w^2+1}+\frac{16}{v
   w^2}+\frac{16}{w^2}+16\right) \\&+ C_F \biggl(-8 v^2 w^4+24 v^2 w^2+\frac{-88 v^2+136 v-48}{-1+v w^2}+\frac{168 v^2+128 v-16}{1+v
   w^2}-16 v^2-4 v w^4\\&+20 v w^2+\frac{-4 w^2-\frac{4}{w^2}+8}{(v-1)^2}+\frac{16}{(v-1) \left(v
   w^2+1\right)}-\frac{64 v}{w^2}+\frac{-4 w^4-8 w^2-\frac{8}{w^2}+12}{v-1}-32 v\\&-4 w^4-4
   w^2-16\biggr) + C_A \left(\frac{-40 v^2-80 v-40}{1+v w^2}+\frac{40 v^2-64 v+24}{-1+v
   w^2}+\frac{8-\frac{8}{w^2}}{v-1}+64\right), \\
c_{\perp,7} = &C_A \left(\frac{16}{v w^4}+\frac{32 (v+1)}{v w^2+1}-\frac{16}{v w^2}-\frac{16}{w^2}-16\right), \\
c_{\perp,8} = &C_F \biggl(-\frac{64 v^2}{w^2}+\frac{-12 v^3-56 v^2+100 v-32}{-1+v w^2}+\frac{76 v^3+24 v^2-84
   v-32}{1+v w^2}\\&+\frac{76 v^3-192 v^2+156 v-40}{\left(-1+v w^2\right)^2}+\frac{12 v^3+32 v^2+28
   v+8}{\left(1+v w^2\right)^2}+\frac{32 v}{w^2}+\frac{8}{w^2}+24\biggr) \\&+ C_A \biggl(\frac{32 v^2}{w^2}+\frac{4 v^3+16 v^2-28 v+8}{-1+vw^2}+\frac{-36 v^3-16 v^2+20 v}{1+v
   w^2}+\frac{-28 v^3+72 v^2-60 v+16}{\left(-1+v w^2\right)^2}\\&+\frac{-12 v^3-32 v^2-28
   v-8}{\left(1+v w^2\right)^2}-\frac{16}{v w^4}+\frac{16}{v w^2}+\frac{16}{w^2}-16\biggr), \\
c_{\perp,9} = &C_F \left(\frac{96 v^2}{w^2}+\frac{-48 v^3+42 v^2+16 v-10}{-1+v w^2}+\frac{-48 v^3-42 v^2+16
   v+10}{1+v w^2}+\frac{28}{w^4}-\frac{36}{w^2}-12\right) \\&+ C_A \left(-\frac{32 v^2}{w^2}+\frac{16 v^3-6 v^2-12 v+2}{-1+v w^2}+\frac{16 v^3+6 v^2-12
   v-2}{1+v w^2}-\frac{12}{w^4}+\frac{8}{w^2}+8\right), \\
c_{\perp,10} = & C_F \biggl(-8 v^3 w^4+24 v^3 w^2-16 v^3+\frac{-20 v^3+60 v^2-52 v+12}{-1+v w^2}+\frac{20 v^3+60
   v^2+52 v+12}{1+v w^2}\\&+\frac{-12 v^3+32 v^2-28 v+8}{\left(-1+v w^2\right)^2}+\frac{-12 v^3-32
   v^2-28 v-8}{\left(1+v w^2\right)^2}+8 v w^2-64 v\biggr) \\&+ C_A \biggl(\frac{-4 v^3+4 v^2-4 v+4}{-1+v w^2}+\frac{4 v^3+4 v^2+4 v+4}{1+v w^2}+\frac{12
   v^3-32 v^2+28 v-8}{\left(-1+v w^2\right)^2}\\&+\frac{12 v^3+32 v^2+28 v+8}{\left(1+v
   w^2\right)^2}+\frac{16}{v w^4}-\frac{32 v}{w^2}-\frac{16}{v w^2}+8 v\biggr), \\
c_{\perp,11} =  & C_F \left(\frac{-88 v^2+64 v+8}{-1+v w^2}+\frac{88 v^2+136 v+48}{1+v w^2}+\frac{16 (v+1)}{v
   w^2-2 v-1}-24\right) \\&+ C_A \left(\frac{40 v^2-64 v+24}{-1+v w^2}-\frac{40 v^2+64 v+24}{1+v w^2}-\frac{8 (v+1)}{v
   w^2-2 v-1}+40\right), \\
c_{\perp,12} =  &C_F \left(-\frac{22 v^2+34 v+12}{1+v w^2}+\frac{22 v^2-34 v-10 w^2+22}{-1+v w^2}-\frac{-4 v-10
   w^2+6}{v w^2-2 v-1}+28\right) \\&+ C_A \left(\frac{10 v^2+16 v+6}{1+v w^2}-\frac{10 v^2-14 v+2 \left(w^2+1\right)}{-1+v
   w^2}+\frac{6 v+2 w^2+4}{v w^2-2 v-1}-4\right).
\end{align*}}
We have checked that the $Q^2 \rightarrow 0 \,\,(w \rightarrow \infty)$ limit of the above agrees with the expression given in~\cite{Ivanov:2004vd, stethesis} for the photoproduction set-up. We recall that 
\begin{equation}
    v = \frac{\hat{\xi}}{w^2}\,\,\,\,\,\,\text{with}\,\,\,\,\,\,w = \sqrt{1-\frac{4m^2}{Q^2}}.
\end{equation}

The bare longitudinal gluon NLO coefficient function can be written in the form 
\begin{equation}
   \hat{C}_{L,g}^{(1)} = \frac{B_1}{N_c (N_c^2-1)}\sqrt{w^2-1}\left( \sum_{i=1}^{12} c_{L,i} f_i + \left\{ v \rightarrow -v \right\} \right),
\end{equation}
where the $f_i$ are given in eqn.~(\ref{fiset}) and the $c_{L,i}$ are
{\allowdisplaybreaks \small
\begin{align*}
c_{L,1} =  &-\frac{C_F}{\epsilon} \frac{12}{w^4} + C_F \biggl(\frac{272 v-256 v^3}{-1+v w^2}+\frac{256 v^3-272 v}{1+v w^2}+\pi ^2 \left(-\frac{64
   v^2}{3 w^2}-\frac{20}{3 w^4}+\frac{28}{3 w^2}\right)\\&+\frac{\pi ^2 \left(\frac{32 v^3}{3}+14
   v^2-\frac{74 v}{3}\right)}{-1+v w^2}+\frac{\pi ^2 \left(\frac{32 v^3}{3}-14 v^2-\frac{74
   v}{3}\right)}{1+v w^2}+\frac{\pi ^2 \left(-\frac{64 v^3}{3}+\frac{128 v^2}{3}-\frac{64
   v}{3}\right)}{\left(-1+v w^2\right)^2}\\&+\frac{256 v^3-512 v^2+256 v}{\left(-1+v
   w^2\right)^2}+\frac{\pi ^2 \left(\frac{64 v^3}{3}+\frac{128 v^2}{3}+\frac{64
   v}{3}\right)}{\left(1+v w^2\right)^2}+\frac{256 v^3+512 v^2+256 v}{\left(1+v
   w^2\right)^2}\\&-\frac{2 w^2}{v-1}+\frac{2 w^2}{v+1}+\frac{2}{(v-1) w^2}-\frac{2}{(v+1)
   w^2}-\frac{16}{w^4}-4 w^2+\frac{32}{w^2}\biggr) \\&+ C_A \biggl(\frac{64 v^3-64 v}{-1+v w^2}+\frac{64 v-64 v^3}{1+v w^2}+\frac{8 \pi ^2
   v^2}{w^2}+\frac{\pi ^2 \left(-4 v^3-\frac{8 v^2}{3}+\frac{20 v}{3}\right)}{-1+v w^2}\\&+\frac{\pi ^2
   \left(-4 v^3+\frac{8 v^2}{3}+\frac{20 v}{3}\right)}{1+v w^2}+\frac{-64 v^3+128 v^2-64 v}{\left(-1+v
   w^2\right)^2}+\frac{\pi ^2 \left(\frac{16 v^3}{3}-\frac{32 v^2}{3}+\frac{16
   v}{3}\right)}{\left(-1+v w^2\right)^2}\\&+\frac{-64 v^3-128 v^2-64 v}{\left(1+v
   w^2\right)^2}+\frac{\pi ^2 \left(-\frac{16 v^3}{3}-\frac{32 v^2}{3}-\frac{16
   v}{3}\right)}{\left(1+v w^2\right)^2}\biggr), \\
c_{L,2} =  &C_F \left(-8 v^2 w^3+32 v^2 w-\frac{160 v^2}{w}+\frac{64 v^3 w-72 v^2 w}{-1+v w^2}+\frac{64
   v^3 w+72 v^2 w}{1+v w^2}-\frac{32}{w^3}+\frac{32}{w}\right)  \\&+ C_A \left(\frac{32 v^2}{w}+\frac{16 v^2 w-16 v^3 w}{-1+v w^2}+\frac{-16 v^3 w-16 v^2 w}{1+v
   w^2}+\frac{16}{w^3}-\frac{8}{w}\right), \\
c_{L,3} = &C_F \left(-\frac{40 v^2}{w^2}+\frac{20 v^3-26 v^2+2 v}{-1+v w^2}+\frac{20 v^3+26 v^2+2 v}{1+v
   w^2}-\frac{8}{w^4}+\frac{12}{w^2}\right) \\&+ C_A \left(\frac{16 v^2}{w^2}+\frac{-8 v^3+16 v^2-8 v}{-1+v w^2}+\frac{-8 v^3-16 v^2-8 v}{1+v
   w^2}\right),\\
c_{L,4} = &C_F \biggl(-8 v^2 w^4+32 v^2 w^2+\frac{-48 v^2+48 v-16}{-1+v w^2}+\frac{80 v^2+80 v}{1+v
   w^2}-32 v^2+4 v w^4-24 v w^2\\&-\frac{16}{(v+1) \left(w^2+1\right)}+\frac{16}{(v+1) \left(-1+v
   w^2\right)}-\frac{16}{\left(w^2+1\right) \left(1+v w^2\right)}-\frac{64
   v+16}{w^2}-\frac{8}{(v+1) w^2}\\&+\frac{4}{(v+1)^2 w^2}+\frac{-4 w^4-16 w^2+44}{v+1}-\frac{-4 w^4+4
   w^2+4}{(v+1)^2}+48 v+20 w^2+\frac{48}{w^2+1}-40\biggr)\\&+C_A \left(\frac{-16 v^2-48 v}{1+v w^2}+\frac{16 v^2+16 v}{-1+v w^2}-\frac{8}{(v+1)
   w^2}+\frac{8}{v+1}+\frac{16}{w^2}-8\right),\\
c_{L,5} = & C_A \left(\frac{16}{v w^4}-\frac{16 v}{-1+v w^2}+\frac{16 v}{1+v w^2}\right), \\
c_{L,6} = &\frac{C_A}{\epsilon}\left(-\frac{16}{v w^4}-\frac{32 v}{1+v w^2}+\frac{16}{w^2}\right) +C_F \biggl(8 v^2 w^4-32 v^2 w^2+\frac{128 v^2-128 v}{-1+v w^2}\\&+\frac{-96 v^2-96 v-32}{1+v
   w^2}+32 v^2+4 v w^4-24 v w^2-\frac{32}{(v-1) \left(1+v w^2\right)}-\frac{64 v}{w^2}\\&+\frac{-4
   w^4-16 w^2-\frac{8}{w^2}+44}{v-1}+\frac{-4 w^4+4 w^2-\frac{4}{w^2}+4}{(v-1)^2}+48 v-20
   w^2+40\biggr)\\& + C_A \biggl(\frac{32 v-32 v^2}{-1+v w^2}+\frac{32 v^2-32 v}{1+v
   w^2}+\frac{8-\frac{8}{w^2}}{v-1}+8\biggr), \\
c_{L,7} = &C_A \left(\frac{16}{v w^4}+\frac{32 v}{1+v w^2}-\frac{16}{w^2}\right), \\
c_{L,8} = &C_F \biggl(-\frac{64 v^2}{w^2}+\frac{96 v^3-16 v^2-80 v}{-1+v w^2}+\frac{-32 v^3-16 v^2+16 v}{1+v
   w^2}+\frac{-128 v^3+256 v^2-128 v}{\left(-1+v w^2\right)^2}\\&+\frac{32
   v}{w^2}+\frac{8}{w^2}\biggr) + C_A \left(\frac{32 v-32 v^3}{-1+v w^2}+\frac{32 v^2}{w^2}+\frac{32 v^3-64 v^2+32 v}{\left(-1+v
   w^2\right)^2}-\frac{16}{v w^4}-\frac{32 v}{1+v w^2}+\frac{16}{w^2}\right), \\
c_{L,9} = &C_F \left(-\frac{48 v^2}{w^2}+\frac{24 v^3-24 v^2}{-1+v w^2}+\frac{24 v^3+24 v^2}{1+v
   w^2}-\frac{8}{w^4}+\frac{8}{w^2}\right) \\&+ C_A \biggl(\frac{16 v^2}{w^2}+\frac{-8 v^3+16 v^2+8 v}{-1+v w^2}+\frac{-8 v^3-16 v^2+8 v}{1+v
   w^2}-\frac{16}{w^2}\biggr), \\ 
c_{L,10} = &C_F \left(8 v^3 w^4-32 v^3 w^2+32 v^3+\frac{16 v^3-32 v^2+16 v}{-1+v w^2}+\frac{-16 v^3-32
   v^2-16 v}{1+v w^2}+16 v\right) \\&+ C_A \left(\frac{-16 v^3+32 v^2-16 v}{-1+v w^2}+\frac{16 v^3+32 v^2+16 v}{1+v w^2}+\frac{16}{v
   w^4}-\frac{32 v}{w^2}-8 v\right), \\
c_{L,11} = &C_F \left(\frac{-128 v^2-128 v}{1+v w^2}+\frac{128 v^2-144 v}{-1+v w^2}\right) + C_A \left(\frac{32 v-32 v^2}{-1+v w^2}+\frac{32 v^2+32 v}{1+v w^2}\right), \\
c_{L,12} = &C_F \left(\frac{-32 v^2+36 v+4}{-1+v w^2}+\frac{32 v^2+32 v}{1+v w^2}-\frac{4 v+4}{v w^2-2
   v-1}\right) \\&+ C_A \left(\frac{8 v^2-16 v+8}{-1+v w^2}-\frac{8 v^2+8 v}{1+v w^2}-\frac{8 (v+1)}{v w^2-2
   v-1}\right).
\end{align*}}
Here, $C_F$ and $C_A$ are the fundamental and adjoint Casimirs of SU($N_c$). With $N_c = 3$, $C_F = 4/3$ and $C_A$ = 3. The diagram group theory factors recur in three combinations: $C_F, C_A$ and $C_F - C_A/2$. The coefficients of $C_F$ and $C_A$ are polynomially reduced and presented above using the \texttt{MultivariateApart} package~\cite{Heller:2021qkz}. The physical symmetry $\hat{\xi} \rightarrow -\hat{\xi}$ is again made apparent in our presentation of the NLO gluon coefficient functions. 

\section{UV renormalisation and mass factorisation}
\label{renorm}

We renormalise the gluon, heavy quark field and heavy quark mass in the on-shell ($\mathrm{OS}$) scheme. The strong coupling constant is renormalised with light flavours treated in the $\overline{\mathrm{MS}}$ scheme and with the heavy quark loop of the gluon self-energy subtracted at zero momentum. The UV renormalised amplitude $\mathcal{A}_\mathrm{UV}$ may be written in terms of the bare amplitude $\mathcal{A}$ using the relation\footnote{Here we omit the renormalisation of the light-quark wave function, which is not relevant at this order as the quark amplitudes enter only at $a_s^2$.},
\begin{align}
    \mathcal{A}^\mathrm{UV} =& Z_A^{n_g/2}  Z_2^{n_q/2} \mathcal{A}(a_0 \rightarrow a_s\, S_\epsilon^{-1} (\mu_R^2/\mu_0^2)^\epsilon Z_\alpha ,\ m_0 \rightarrow m Z_{m} ) \nonumber \\
    =& 
    a_s\, S_\epsilon^{-1} \left(\frac{\mu_R^2}{\mu_0^2}\right)^\epsilon \biggl( \mathcal{A}^{(0)}
    + a_s\, \left( \frac{n_g}{2} \delta Z_A + \frac{n_q}{2} \delta Z_2 + \delta Z_\alpha \right) \mathcal{A}^{(0)} \nonumber \\
    & + a_s\, \delta Z_{m} \mathcal{A}^{\mathrm{mct},(0)} + 
    a_s\, S_\epsilon^{-1} \left(\frac{\mu_R^2}{\mu_0^2}\right)^{\epsilon} \mathcal{A}^{(1)} + \mathcal{O}(a_s^2) \biggr).
\end{align}
Here $a_0=\alpha_0/(4\pi)$ and $ a_s = \alpha_s(\mu_R^2)/(4\pi)$ are the bare and renormalised strong couplings, respectively, and we have introduced $S_\epsilon = (4 \pi)^\epsilon e^{-\epsilon \gamma_E}$.
The heavy quark bare mass parameter is denoted $m_0$ and the renormalised quark mass is denoted $m$. The gluon and heavy quark renormalisation constants are denoted $Z_A$ and $Z_2$ and the number of external gluons and heavy quarks are denoted $n_g$ and $n_q$, respectively. In the second line the bare amplitude and renormalisation constants are expanded using
\begin{align}
\mathcal{A} &= a_0 \mathcal{A}^{(0)} + a_0^2 \mathcal{A}^{(1)} + \mathcal{O}(a_0^3), \nonumber \\
Z_i &= 1 + a_s \delta Z_i + \mathcal{O}(a_s^2) \qquad (i=A,2,\alpha,m).
\end{align}
We have also introduced the mass counterterm amplitude $\mathcal{A}^\mathrm{mct}$ which is computed by inserting mass counter vertices and cross-checked by computing the derivative of the bare amplitude with respect to the heavy quark mass, $m$. The derivative must be taken prior to equating the HVM mass, $M$, to twice the heavy quark mass in the HVM projector. 

The explicit expressions for the renormalisation constants at one-loop are
\begin{align}
\delta Z_\alpha &= -\frac{1}{\epsilon} \beta_0 + \delta Z_\alpha^\mathrm{hq},\qquad \beta_0 = \frac{11}{3} C_A - \frac{4}{3} T_F\, n_f, \\
\delta Z_A &= - \delta Z_\alpha^\mathrm{hq} = \left(\frac{\mu_R^2}{m^2}\right)^\epsilon \left( -\frac{4}{3\epsilon} T_F \right), \\
\delta Z_2 &=  \delta Z_{m} = \left(\frac{\mu_R^2}{m^2}\right)^\epsilon C_F \left( -\frac{3}{\epsilon} - 4 \right). 
\end{align}
Although the above renormalisation procedure is stated for the bare amplitudes, it can equally be applied to each of the bare coefficient functions $\hat{C}_{i=\perp,L,\, j=g,q}$. The UV renormalised coefficient functions are thus given by
\begin{align}
\hat{C}_{i,j}^{(0),\mathrm{UV}} = & S_\epsilon^{-1} \left(\frac{\mu_R^2}{\mu_0^2}\right)^\epsilon \hat{C}_{i,j}^{(0)}, \\ \hat{C}^{(1),\mathrm{UV}}_{i,j} = & S_\epsilon^{-2} \left(\frac{\mu_R^2}{\mu_0^2}\right)^{2 \epsilon} \hat{C}_{i,j}^{(1)} +  \left( \frac{n_g}{2} \delta Z_A + \frac{n_q}{2} \delta Z_2 + \delta Z_\alpha \right) \hat{C}_{i,j}^{(0),\mathrm{UV}} \nonumber \\ 
& + S_\epsilon^{-1} \left(\frac{\mu_R^2}{\mu_0^2}\right)^\epsilon \delta Z_{m} \hat{C}_{i,j}^{\mathrm{mct},(0)}.
\end{align}
The relevant mass counterterm coefficient functions are
\begin{align}
\hat{C}_{\perp,g}^{\mathrm{mct},(0)} & = \frac{4 B_0}{N_c (N_c^2-1)} \biggl( \frac{v-1}{v w^2-1}(1-\epsilon)+\frac{v+1}{v w^2+1}(1-\epsilon)-\frac{2}{w^4}+\frac{3}{w^2}-3 + 2\epsilon \biggr), \\
\hat{C}_{L,g}^{\mathrm{mct},(0)} & = \frac{8B_0}{N_c (N_c^2-1)} \frac{(w^2-1)^{3/2}}{w^4}.
\end{align}
Since the quark amplitudes enter only at $a_s^2$ they are trivially modified by the above procedure.

After UV renormalisation, poles in $\epsilon$ still remain in both the quark and gluon one-loop coefficient functions and must be absorbed into the definition of the GPDs via mass factorisation. This procedure generates counterterms which render the coefficient functions finite. Concretely, we replace the bare quark singlet ($\hat{F}^S$) and gluon ($\hat{F}^g$) GPDs with the mass factorised, factorisation scale ($\mu_F^2$) dependent, GPDs,
\begin{align}
\begin{split}
\hat{F}^S(x,\xi) &= F^S(x,\xi,\mu_F^2) + \frac{a_s}{\epsilon} \left( \frac{\mu_R^2}{\mu_F^2} \right)^\epsilon \\
&\times \int_{-1}^1 \frac{\text{d} z}{| \xi |} \left[ V^{(1)}_{qq} \left( \frac{x}{\xi},\frac{z}{\xi} \right) F^S(z, \xi, \mu_F^2) + \xi^{-1} V^{(1)}_{qg} \left(\frac{x}{\xi},\frac{z}{\xi} \right) F^g(z, \xi, \mu_F^2) \right], \\
\hat{F}^g(x,\xi) &= F^g(x,\xi,\mu_F^2) + \frac{a_s}{\epsilon} \left( \frac{\mu_R^2}{\mu_F^2} \right)^\epsilon \\
&\times \int_{-1}^1 \frac{\text{d} z}{| \xi |}  \left[ \xi V^{(1)}_{gq} \left( \frac{x}{\xi},\frac{z}{\xi} \right) F^S(z, \xi, \mu_F^2) + V^{(1)}_{gg} \left(\frac{x}{\xi},\frac{z}{\xi} \right) F^g(z, \xi, \mu_F^2) \right]. \label{eq:GPDBareToRen}
\end{split}
\end{align}
Here $V^{(1)}$ is the coefficient of $\alpha_s/(4 \pi)$ in the generalised splitting function $V$. We take the generalised splitting functions from Ref.~\cite{Diehl:2007hd}.
Inserting these relations into the bare version of the factorisation formula \eqref{eq:HVMPartonFac} and absorbing the divergent terms into the coefficient functions we obtain
\begin{align}
    C_{i,j}^{(1)} &= \hat{C}_{i,j}^{(1),\mathrm{UV}} + \hat{C}_{i,j}^{\mathrm{mf},(1)},
\end{align}
where the mass factorisation counterterms are given by
\begin{align}
\hat{C}_{\perp,g}^{\mathrm{mf},(1)} & = 
\frac{1}{\epsilon} \left( \frac{\mu_R^2}{\mu_F^2} \right)^\epsilon \frac{1}{|\xi|} \int_{-1}^1 \frac{\mathrm{d} z}{z^2} \hat{C}_{\perp,g}^{(0),\text{UV}}\left(\frac{\xi}{z},Q^2\right) V^{(1)}_{gg} \left(\frac{z}{\xi},\frac{x}{\xi} \right) \nonumber \\ 
& = \frac{1}{\epsilon} \left( \frac{\mu_R^2}{\mu_F^2} \right)^{\epsilon} S_\epsilon^{-1} \left( \frac{\mu_R^2}{\mu_0^2} \right)^{\epsilon}  \frac{4\,T_F}{N_c} B_0\biggl( C_A \left(\frac{1}{v w^4}+\frac{2 v+2}{v w^2+1}-\frac{1}{v w^2}-\frac{1}{w^2}-1\right) f_6 \nonumber \\&\,\,\,\,\,\,\,\,\,\,\,\,\,\,\,\,\,\,\,\,\,\,\,\,\,\,\,\,\,\,\,\,\,\,\,\,\,\,\,\,\,\,\,\,\,\,\,\,\,\,\,\,\,\,\,\,\,\,\,\,\,\,\,\,\,\,\,\,\,\,\,\,\,\,\,\,\,\,\,\,\,\,\,\,\,\,\,\,\, +\frac{\beta_0}{4} \frac{w^2-1}{w^2} + \left\{v \rightarrow -v\right\}\biggr), \\ \nonumber \\
\hat{C}_{L,g}^{\mathrm{mf},(1)}  & = 
\frac{1}{\epsilon} \left( \frac{\mu_R^2}{\mu_F^2} \right)^\epsilon \frac{1}{|\xi|} \int_{-1}^1 \frac{\mathrm{d} z}{z^2} \hat{C}_{L,g}^{(0),\text{UV}}\left(\frac{\xi}{z}, Q^2\right) V^{(1)}_{gg} \left(\frac{z}{\xi},\frac{x}{\xi} \right) 
 = - \frac{\hat{C}_{\perp,g}^{\mathrm{mf},(1)}}{\sqrt{w^2-1}}, \\ \nonumber \\
\hat{C}_{\perp,q}^{\mathrm{mf},(1)} & = 
\frac{1}{\epsilon} \left( \frac{\mu_R^2}{\mu_F^2} \right)^\epsilon  \frac{\xi}{|\xi|} \int_{-1}^1 \frac{\mathrm{d} z}{z^2} \hat{C}_{\perp,g}^{(0),\text{UV}}\left(\frac{\xi}{z},Q^2\right) 2V^{(1)}_{gq} \left(\frac{z}{\xi},\frac{x}{\xi} \right) \nonumber \\
& =  \frac{1}{\epsilon} \left( \frac{\mu_R^2}{\mu_F^2} \right)^{\epsilon} S_\epsilon^{-1} \left( \frac{\mu_R^2}{\mu_0^2} \right)^{\epsilon} \frac{64\, T_F^2}{N_c^2} B_0 \frac{x_+ x_-}{x^2} \left(\frac{1}{v w^2}+\frac{1}{w^2}-\frac{1}{v w^4}-\frac{v+1}{v w^2+1}\right)f_6 \nonumber \\&\,\,\,\,\,\,\,\,\,\,\,\,\,\,\,\,\,\,\,\,\,\,\,\,\,\,\,\,\,\,\,\,\,\,\,\,\,\,\,\,\,\,\,\,\,\,\,\,\,\,\,\,\,\,\,\,\,\,\,\,\,\,\,\,\,\,\,\,\,\,\,\,\,\,\,\,\,\,\,\,\,\,\,\,\,\,\,\,\,\,\,\,\,\,\,\,\,\,\,\,\,\,\,\,\,+ \left\{v \rightarrow -v\right\}, \\ \nonumber \\
\hat{C}_{L,q}^{\mathrm{mf},(1)} & = 
\frac{1}{\epsilon} \left( \frac{\mu_R^2}{\mu_F^2} \right)^\epsilon \frac{\xi}{|\xi|} \int_{-1}^1 \frac{\mathrm{d} z}{z^2} \hat{C}_{L,g}^{(0),\text{UV}}\left(\frac{\xi}{z}, Q^2\right) 2V^{(1)}_{gq} \left(\frac{z}{\xi},\frac{x}{\xi} \right) = - \frac{\hat{C}_{\perp,q}^{\mathrm{mf},(1)}}{\sqrt{w^2-1}}.
\end{align}
The final renormalised coefficient functions are available in an ancillary \texttt{MATHEMATICA} file alongside the arXiv entry of this paper.

\section{Checks on our results and comparison with literature}

We have performed the following checks on our results. As stated already, the $Q^2 \rightarrow 0$ limit of our electroproduction result coincides with the photoproduction result already available in the literature~\cite{Ivanov:2004vd, stethesis}. In this limit, the longitudinal component vanishes and there is only a non-vanishing amplitude to produce a transversely polarised HVM from a transversely polarised photon. Note that the $m^2 \rightarrow 0$ limit is not smooth and cannot be taken. Our results are written with the symmetry $\xi \rightarrow -\xi$ manifest. The tensor decomposition of the amplitude in eqn.~(\ref{tensordecomp1}) was verified explicitly and all partial fraction linear reductions were checked numerically. Moreover, at one-loop, gluon propagators introduce a gauge parameter dependence on the Feynman diagram level if computed in a linear covariant gauge, as done here. The cancellation of all gauge parameter dependent terms at the amplitude level provides a further check. 

The authors of~\cite{recent} have also computed the transverse and longitudinal renormalised amplitudes at NLO for the exclusive electroproduction of heavy quarkonia, within the same calculational framework of collinear factorisation and  LO NRQCD. Our work therefore serves as an independent, simultaneous analysis of the process and, in addition, a check on the results already available.  We have obtained agreement numerically with both their quark and gluon transverse and longitudinal renormalised amplitudes after publication of their erratum, in which typographical errors were corrected and an inconsistency resolved in their high energy limit ($\xi \rightarrow 0$) that we had pointed out. Interestingly, in the high energy limit the dominant logarithm $\ln(1/\xi)$ is multiplied by $\ln \left(\bar{Q}^2/\mu_F^2 \right)$ where $\bar{Q}^2 = (Q^2+4m^2)/4$. 
Therefore, adopting the scale choice $\mu_F^2 = \bar{Q}^2$ would automatically resum these double-logarithmic terms $\sim\alpha_s^2 \ln \left(\bar{Q}^2/\mu_F^2 \right) \ln(1/\xi)$ and effectively set the NLO corrections to zero in the high energy limit. 

Note that our results are expressed in terms of a basis set comprising a smaller number of logarithms and dilogarithms. 
The set we have selected is such that all basis functions appearing are manifestly real {\it if} the coefficient function is real. 
Throughout the entire ERBL phase space, all of our basis functions evaluate to a real number and, in addition, only one square root argument is not rationalised. The substitution $z^2 = (1+2v-vw^2)/(1+vw^2)$ would allow all arguments of the functions appearing in our basis set to be rationalised, however this would upset our explicit $v\rightarrow -v$ symmetry which we deem to be a more important feature to maintain. 
We have also decided to keep track of all the group theory factors that arise, allowing us to generalise the expressions for $N_c \neq 3$ and be valid for an arbitrary SU($N_c$) gauge theory instead.

\section{Conclusions}
In this work, we have computed renormalised coefficient functions for exclusive electroproduction of heavy vector mesons to NLO in the collinear factorisation framework. The description of the $Q \bar{Q} \rightarrow V$ transition vertex was made within LO NRQCD and the mass of the heavy vector meson set equal to twice the on-shell mass of the heavy quark. Our results respect the physical $\xi \rightarrow -\xi$ symmetry and coincide with the renormalised coefficient functions for exclusive photoproduction of heavy vector mesons in the $Q^2 \rightarrow 0$ limit.

They can be used in a phenomenological analysis of the exclusive electroproduction data already measured at HERA and, in time, with the data from LHC, the upcoming EIC and the proposed LHeC and FCC. 
These results may also help in studying the onset of saturation physics.
Importantly, our predictions and these data can collectively provide further constraints on the gluon distribution in nuclei at moderate to low values of the scale $Q^2$ and Bjorken $x$.

\acknowledgments

We would like to thank Zi-Qiang Chen and Cong-Feng Qiao for discussions and for communicating their erratum prior to its publication. CAF is supported by the Helsinki Institute of Physics core funding project QCD-THEORY (project 7915122). The work of CAF was funded in part through an STFC PhD
studentship (Grant ST/N504130/1). JAG and TT are supported by STFC Consolidated Grants ST/L000431/1, ST/T000988/1 and ST/P000290/1. The work of JAG was also supported in part by a DFG Mercator Fellowship. SPJ is supported by a Royal Society University Research Fellowship (Grant URF/R1/201268). All Figures were produced using JaxoDraw~\cite{BINOSI200476}.

\clearpage

\bibliographystyle{jhep}
\bibliography{references.bib}

\providecommand{\href}[2]{#2}\begingroup\raggedright\begin{thebibliography}{10}

\bibitem{Ryskin:1992ui}
M.~Ryskin, {\it {Diffractive $J /\psi$ electroproduction in LLA QCD}},  {\em Z.
  Phys. C} {\bf 57} (1993) 89--92.

\bibitem{Ivanov:2004ax}
I.~Ivanov, N.~Nikolaev, and A.~Savin, {\it {Diffractive vector meson production
  at HERA: From soft to hard QCD}},  {\em Phys. Part. Nucl.} {\bf 37} (2006)
  1--85, [\href{http://arxiv.org/abs/hep-ph/0501034}{{\tt hep-ph/0501034}}].

\bibitem{Collins:1996fb}
J.~C. Collins, L.~Frankfurt, and M.~Strikman, {\it {Factorization for hard
  exclusive electroproduction of mesons in QCD}},  {\em Phys. Rev. D} {\bf 56}
  (1997) 2982--3006, [\href{http://arxiv.org/abs/hep-ph/9611433}{{\tt
  hep-ph/9611433}}].

\bibitem{Augustyniak:2013ypa}
{\bf HERMES} Collaboration, W.~Augustyniak, {\it {Exclusive electroproduction
  of vector mesons in lepton nucleon scattering at the HERMES experiment}},
  {\em Nucl. Phys. B Proc. Suppl.} {\bf 245} (2013) 207--214.

\bibitem{Chekanov:2004mw}
{\bf ZEUS} Collaboration, S.~Chekanov et~al., {\it {Exclusive electroproduction
  of $J/\psi$ mesons at HERA}},  {\em Nucl. Phys. B} {\bf 695} (2004) 3--37,
  [\href{http://arxiv.org/abs/hep-ex/0404008}{{\tt hep-ex/0404008}}].

\bibitem{Aktas:2005xu}
{\bf H1} Collaboration, A.~Aktas et~al., {\it {Elastic $J/\psi$ production at
  HERA}},  {\em Eur. Phys. J. C} {\bf 46} (2006) 585--603,
  [\href{http://arxiv.org/abs/hep-ex/0510016}{{\tt hep-ex/0510016}}].

\bibitem{Aaij:2014iea}
{\bf LHCb} Collaboration, R.~Aaij et~al., {\it {Updated measurements of
  exclusive $J/\psi$ and $\psi$(2S) production cross-sections in pp collisions
  at $\sqrt{s}=7$ TeV}},  {\em J. Phys. G} {\bf 41} (2014) 055002,
  [\href{http://arxiv.org/abs/1401.3288}{{\tt arXiv:1401.3288}}].

\bibitem{Aaij:2015kea}
{\bf LHCb} Collaboration, R.~Aaij et~al., {\it {Measurement of the exclusive
  $\Upsilon$ production cross-section in pp collisions at $ \sqrt{s}=7 $ TeV
  and 8 TeV}},  {\em JHEP} {\bf 09} (2015) 084,
  [\href{http://arxiv.org/abs/1505.08139}{{\tt arXiv:1505.08139}}].

\bibitem{Aaij:2018arx}
{\bf LHCb} Collaboration, R.~Aaij et~al., {\it {Central exclusive production of
  $J/\psi$ and $\psi(2S)$ mesons in $pp$ collisions at $\sqrt{s}=13~$TeV}},
  {\em JHEP} {\bf 10} (2018) 167, [\href{http://arxiv.org/abs/1806.04079}{{\tt
  arXiv:1806.04079}}].

\bibitem{Jones:2016icr}
S.~P. Jones, A.~D. Martin, M.~G. Ryskin, and T.~Teubner, {\it {Exclusive
  $J/\psi$ production at the LHC in the $k_T$ factorization approach}},  {\em
  J. Phys. G} {\bf 44} (2017), no.~3 03LT01,
  [\href{http://arxiv.org/abs/1611.03711}{{\tt arXiv:1611.03711}}].

\bibitem{Ivanov:2004vd}
D.~Ivanov, A.~Schafer, L.~Szymanowski, and G.~Krasnikov, {\it {Exclusive
  photoproduction of a heavy vector meson in QCD}},  {\em Eur. Phys. J. C} {\bf
  34} (2004), no.~3 297--316, [\href{http://arxiv.org/abs/hep-ph/0401131}{{\tt
  hep-ph/0401131}}]. [Erratum: Eur. Phys. J. C 75, 75 (2015)].

\bibitem{stethesis}
S.~P. Jones, {\it {A Study of Exclusive Processes to NLO and Small $x$ PDFs
  from LHC data}},  The University of Liverpool, 2014, {\it Unpublished}.

\bibitem{Ivanov:2004pp}
D.~Ivanov, M.~Kotsky, and A.~Papa, {\it {The Impact factor for the virtual
  photon to light vector meson transition}},  {\em Eur. Phys. J. C} {\bf 38}
  (2004) 195--213, [\href{http://arxiv.org/abs/hep-ph/0405297}{{\tt
  hep-ph/0405297}}].

\bibitem{Jones:2015nna}
S.~P. Jones, A.~D. Martin, M.~G. Ryskin, and T.~Teubner, {\it {Exclusive
  $J/\psi$ and $\Upsilon$ photoproduction and the low $x$ gluon}},  {\em J.
  Phys. G} {\bf 43} (2016), no.~3 035002,
  [\href{http://arxiv.org/abs/1507.06942}{{\tt arXiv:1507.06942}}].

\bibitem{Jones:2016ldq}
S.~P. Jones, A.~D. Martin, M.~G. Ryskin, and T.~Teubner, {\it {The exclusive
  $J/\psi$ process at the LHC tamed to probe the low $x$ gluon}},  {\em Eur.
  Phys. J. C} {\bf 76} (2016), no.~11 633,
  [\href{http://arxiv.org/abs/1610.02272}{{\tt arXiv:1610.02272}}].

\bibitem{Flett:2019pux}
C.~A. Flett, S.~P. Jones, A.~D. Martin, M.~G. Ryskin, and T.~Teubner, {\it {How
  to include exclusive $J/\psi$ production data in global PDF analyses}},  {\em
  Phys. Rev. D} {\bf 101} (2020), no.~9 094011,
  [\href{http://arxiv.org/abs/1908.08398}{{\tt arXiv:1908.08398}}].

\bibitem{Flett:2020duk}
C.~A. Flett, A.~D. Martin, M.~G. Ryskin, and T.~Teubner, {\it {Very low $x$
  gluon density determined by LHCb exclusive $J/\psi$ data}},  {\em Phys. Rev.
  D} {\bf 102} (2020) 114021, [\href{http://arxiv.org/abs/2006.13857}{{\tt
  arXiv:2006.13857}}].

\bibitem{AbdulKhalek:2021gbh}
R.~Abdul~Khalek et~al., {\it {Science Requirements and Detector Concepts for
  the Electron-Ion Collider: EIC Yellow Report}},
  \href{http://arxiv.org/abs/2103.05419}{{\tt arXiv:2103.05419}}.

\bibitem{LHeC:2020van}
{\bf LHeC, FCC-he Study Group} Collaboration, P.~Agostini et~al., {\it {The
  Large Hadron-Electron Collider at the HL-LHC}},
  \href{http://arxiv.org/abs/2007.14491}{{\tt arXiv:2007.14491}}.

\bibitem{FCC:2018byv}
{\bf FCC} Collaboration, A.~Abada et~al., {\it {FCC Physics Opportunities}:
  {Future Circular Collider Conceptual Design Report Volume 1}},  {\em Eur.
  Phys. J. C} {\bf 79} (2019), no.~6 474.

\bibitem{Hoodbhoy:1996zg}
P.~Hoodbhoy, {\it {Wave function corrections and off forward gluon
  distributions in diffractive $J / \psi$ electroproduction}},  {\em Phys. Rev.
  D} {\bf 56} (1997) 388--393, [\href{http://arxiv.org/abs/hep-ph/9611207}{{\tt
  hep-ph/9611207}}].

\bibitem{Ji:1998xh}
X.-D. Ji and J.~Osborne, {\it {One loop corrections and all order factorization
  in deeply virtual Compton scattering}},  {\em Phys. Rev. D} {\bf 58} (1998)
  094018, [\href{http://arxiv.org/abs/hep-ph/9801260}{{\tt hep-ph/9801260}}].

\bibitem{Petrelli:1997ge}
A.~Petrelli, M.~Cacciari, M.~Greco, F.~Maltoni, and M.~L. Mangano, {\it {NLO
  production and decay of quarkonium}},  {\em Nucl. Phys. B} {\bf 514} (1998)
  245--309, [\href{http://arxiv.org/abs/hep-ph/9707223}{{\tt hep-ph/9707223}}].

\bibitem{Bodwin:2013zu}
G.~T. Bodwin and A.~Petrelli, {\it {Order-$v^4$ corrections to $S$-wave
  quarkonium decay}},  {\em Phys. Rev. D} {\bf 66} (2002) 094011,
  [\href{http://arxiv.org/abs/hep-ph/0205210}{{\tt hep-ph/0205210}}]. [Erratum:
  Phys. Rev. D 87, 039902 (2013)].

\bibitem{Braaten:2002fi}
E.~Braaten and J.~Lee, {\it {Exclusive Double Charmonium Production from $e^+
  e^-$ Annihilation into a Virtual Photon}},  {\em Phys. Rev. D} {\bf 67}
  (2003) 054007, [\href{http://arxiv.org/abs/hep-ph/0211085}{{\tt
  hep-ph/0211085}}]. [Erratum: Phys. Rev. D 72, 099901 (2005)].

\bibitem{Braun:2002wu}
V.~M. Braun, D.~Ivanov, A.~Schafer, and L.~Szymanowski, {\it {Towards the
  theory of coherent hard dijet production on hadrons and nuclei}},  {\em Nucl.
  Phys. B} {\bf 638} (2002) 111--154,
  [\href{http://arxiv.org/abs/hep-ph/0204191}{{\tt hep-ph/0204191}}].

\bibitem{Pire:2011st}
B.~Pire, L.~Szymanowski, and J.~Wagner, {\it {NLO corrections to timelike,
  spacelike and double deeply virtual Compton scattering}},  {\em Phys. Rev. D}
  {\bf 83} (2011) 034009, [\href{http://arxiv.org/abs/1101.0555}{{\tt
  arXiv:1101.0555}}].

\bibitem{Belitsky:2012ch}
A.~V. Belitsky, D.~Müller, and Y.~Ji, {\it {Compton scattering: from deeply
  virtual to quasi-real}},  {\em Nucl. Phys. B} {\bf 878} (2014) 214--268,
  [\href{http://arxiv.org/abs/1212.6674}{{\tt arXiv:1212.6674}}].

\bibitem{Nogueira:1991ex}
P.~Nogueira, {\it {Automatic Feynman graph generation}},  {\em J. Comput.
  Phys.} {\bf 105} (1993) 279--289.

\bibitem{Ruijl:2017dtg}
B.~Ruijl, T.~Ueda, and J.~Vermaseren, {\it {FORM version 4.2}},
  \href{http://arxiv.org/abs/1707.06453}{{\tt arXiv:1707.06453}}.

\bibitem{Lei78}
E.~Leinartas, {\it {Factorization of rational functions of several variables
  into partial fractions}},  {\em Soviet Math. (Iz. VUZ)} {\bf 22} (1978)
  35--38.

\bibitem{vonManteuffel:2012np}
A.~von Manteuffel and C.~Studerus, {\it {Reduze 2 - Distributed Feynman
  Integral Reduction}},  \href{http://arxiv.org/abs/1201.4330}{{\tt
  arXiv:1201.4330}}.

\bibitem{Laporta:2001dd}
S.~Laporta, {\it {High precision calculation of multiloop Feynman integrals by
  difference equations}},  {\em Int. J. Mod. Phys. A} {\bf 15} (2000)
  5087--5159, [\href{http://arxiv.org/abs/hep-ph/0102033}{{\tt
  hep-ph/0102033}}].

\bibitem{Heller:2021qkz}
M.~Heller and A.~von Manteuffel, {\it {MultivariateApart: Generalized Partial
  Fractions}},  \href{http://arxiv.org/abs/2101.08283}{{\tt arXiv:2101.08283}}.

\bibitem{Diehl:2007hd}
M.~Diehl and W.~Kugler, {\it {Next-to-leading order corrections in exclusive
  meson production}},  {\em Eur. Phys. J. C} {\bf 52} (2007) 933--966,
  [\href{http://arxiv.org/abs/0708.1121}{{\tt arXiv:0708.1121}}].

\bibitem{recent}
Z.-Q. Chen and C.-F. Qiao, {\it {NLO QCD corrections to exclusive
  electroproduction of quarkonium}},  {\em Phys. Lett. B} {\bf 797} (2019)
  134816, [\href{http://arxiv.org/abs/1903.00171}{{\tt arXiv:1903.00171}}].
  [Erratum: Phys. Lett. B, 135759 (2020)].

\bibitem{BINOSI200476}
D.~Binosi and L.~Theußl, {\it Jaxodraw: A graphical user interface for drawing
  feynman diagrams},  {\em Comput. Phys. Commun.} {\bf 161} (2004), no.~1
  76--86.

\end{thebibliography}\endgroup

\end{document}